\documentstyle[11pt,aaspp4]{article}

\singlespace

\def\Kp{\ifmmode{K^{\prime}}\else{${K^{\prime}}$}\fi}
\def\MKp{\ifmmode{\overline M_{K^{\prime}}}\else{$\overline M_{K^{\prime}}$}\fi}
\def\mKp{\ifmmode{\overline m_{K^{\prime}}}\else{$\overline m_{K^{\prime}}$}\fi}
\def\MK{\ifmmode{\overline M_{K}}\else{$\overline M_{K}$}\fi}
\def\mK{\ifmmode{\overline m_{K}}\else{$\overline m_{K}$}\fi}
\def\MI{\ifmmode{\overline M_{I}}\else{$\overline M_{I}$}\fi}
\def\mI{\ifmmode{\overline m_{I}}\else{$\overline m_{I}$}\fi}
\def\MH{\ifmmode{\overline M_{H}}\else{$\overline M_{H}$}\fi}
\def\mH{\ifmmode{\overline m_{H}}\else{$\overline m_{H}$}\fi}
\def\MJ{\ifmmode{\overline M_{J}}\else{$\overline M_{J}$}\fi}
\def\mJ{\ifmmode{\overline m_{J}}\else{$\overline m_{J}$}\fi}
\def\p0/p1{\ifmmode{P_0/P_1}\else{$P_0/P_1$}\fi}
\def\snsbf{\ifmmode{\xi}\else{$\xi$}\fi}
\def\E{\ifmmode{E(k)}\else{$E(k)$}\fi}
\def\mg2{Mg$_2$}
\def\Mg2{Mg$_2$}
\def\v-i{\ifmmode{(V{-}I)}\else{$(V{-}I)$}\fi}
\def\m-M{\ifmmode{(m{-}M)}\else{$(m{-}M)$}\fi}
\def\m-Mbar{\ifmmode{(\overline m{-}\overline M)}\else{$(\overline m{-}\overline M)$}\fi}
\def\kms{\ifmmode{{\rm km\,s}^{-1}}\else{km\,s$^{-1}$}\fi}

\def\eps{\ifmmode{{\rm e}^-\,{\rm s}^{-1}}\else{e$^-$\,s$^{-1}$}\fi}
\def\epspp{\ifmmode{{\rm e}^-\,{\rm s}^{-1}\,{\rm pixel}^{-1}}\else{e$^-$\,s$^{-1}$\,pixel$^{-1}$}\fi}

\begin{document}

\title{Measuring Distances Using Infrared Surface Brightness Fluctuations}

\author{Joseph B. Jensen\altaffilmark{1}}
\author{John L. Tonry}
\and
\author{Gerard A. Luppino}
\affil{Institute for Astronomy, University of Hawaii\\
	2680 Woodlawn Drive, Honolulu, HI  96822\\
	e-mail: jjensen@gemini.edu,
	jt@avidya.ifa.hawaii.edu,
	ger@hokupa.ifa.hawaii.edu} 
\altaffiltext{1}{Currently with the Gemini 8 m Telescopes Project,
	180 Kinoole St. Suite 207, Hilo, HI  96720.  The Gemini 8 m  
	Telescopes Project is managed by the Association of 
	Universities for Research in Astronomy, for the National Science 
	Foundation and the Gemini Board, under an international 
	partnership agreement.}
\authoraddr{2680 Woodlawn Drive, Honolulu, HI  96822}

\slugcomment{Accepted for publication in {\it The Astrophysical Journal}}

\begin{abstract}

Surface brightness fluctuations (SBFs) are much brighter in the infrared
than they are at optical wavelengths, 
making it possible to measure greater
distances using IR SBFs. 
We report new \Kp\ (2.1\micron) SBF measurements
of nine galaxies in the Fornax and Eridanus clusters 
using a 1024$^2$-pixel HgCdTe array.
We used improved analysis techniques to remove contributions to the SBFs
from globular clusters and background galaxies, 
and we assess the relative importance of other sources of residual variance. 
We applied the improved methodology to our Fornax and Eridanus images
and to our previously published Virgo cluster data.
Apparent fluctuation magnitudes were used in conjunction with 
Cepheid distances to M31 and the Virgo cluster to calibrate the \Kp\ 
SBF distance scale.  
We find the absolute fluctuation magnitude $\MKp\,{=}\,{-}5.61\,{\pm}\,0.12$,
with an intrinsic scatter to the calibration of 0.06 mag. 
No statistically significant change in \MKp\ is detected as
a function of \v-i.  
Our calibration is consistent with simple (constant age and metallicity)
stellar population models.
The lack of a correlation with \v-i\ in the context of the stellar population
models implies that elliptical galaxies bluer
than $\v-i\,{=}\,1.2$ have SBFs dominated by younger (5--8 Gyr) populations
and metallicities comparable to redder ellipticals.
Significant contributions to the SBFs from anomalous populations of 
asymptotic giant branch stars
are apparently uncommon in giant ellipticals.
\Kp\ SBFs prove to be a reliable distance indicator as long as the 
residual variance from globular clusters and background galaxies
is properly removed.  Also, it is important that a sufficiently high
signal-to-noise ratio be achieved to allow reliable sky subtraction
because residual spatial variance can bias the measurement
of the SBF power spectrum.
\end{abstract}


\section{Introduction}

Because the light from a distant galaxy comes from discrete but
unresolved stars,
Poisson statistics lead to mottling of the galaxy's otherwise smooth 
surface brightness profile.
Surface brightness is independent of distance, but the amplitude
of the surface brightness fluctuations (SBFs) is not.
As the distance ($d$) to a galaxy increases, the number of stars ($n$) 
in a given resolution element increases as $d^2$, but the observed flux ($f$) 
from each star is reduced by $d^{-2}$, 
making surface brightness ($nf$) independent of distance.
On the other hand, the rms variation in observed flux from region to 
region is $n^{1/2}f$, which scales as $d^{-1}$.  The variance in
surface brightness $nf^2$ normalized by the mean galaxy brightness $nf$
decreases with distance; distant galaxies appear smoother than
nearby galaxies. 
Because the variance in surface brightness is
proportional to the second moment of the stellar luminosity function,
it is dominated by luminous red giant stars.

Although individual stars are not resolved, measuring SBFs probes 
the stellar population of the galaxy directly.
With good theoretical models for stellar populations, the absolute
magnitude of SBFs can be calculated, allowing a direct determination
of the distance that is independent of the global dynamics
or the environment of the galaxy (to the extent that the stellar populations
in old stellar systems are independent of these variables).
Alternatively, measuring SBFs in 
galaxies with known distances provides insight into stellar populations,
allowing comparison
with stellar evolution models and providing an empirical calibration
of the SBF distance scale.
Good descriptions of the theory and practice of using SBFs as a 
distance measurement tool and stellar
population probe can be found in several papers by Tonry and 
collaborators (Tonry et al. 1997; Jensen, Luppino, \& Tonry 1996,
hereafter JLT;
Tonry 1991; Tonry, Ajhar, \& Luppino 1990; Tonry \& Schneider 1988),
in papers by Sodemann \& Thomsen (1995, 1996),
and in the review by Jacoby et al. (1992).
The first $K$-band SBF 
studies are described by JLT,
Luppino \& Tonry (1993), and Pahre \& Mould (1994).

J. Tonry and coworkers have completed an extensive 
survey of $I$-band SBF distances in a sample of  
several hundred early-type (dynamically hot elliptical and S0) 
galaxies which is more than 50\% complete
out to $\sim$2800 \kms\ (Tonry et al. 1997).
They find that the $I$-band absolute fluctuation magnitude
\MI\ is a linear function of \v-i\ and has a universal zero point.
Their $I$-band calibration, which is based empirically on Cepheid distances,
is in good agreement with Worthey's (1993a,b; 1994) 
simple stellar populations models.
Worthey's models predict that the effects of age and metallicity in the $I$
band are largely degenerate, so that \MI\ may be calibrated using a single
parameter such as the \v-i\ color.  
The resulting intrinsic scatter of the $I$-band SBF distance scale is of
order 0.07 mag.  
The purpose of the current study is to examine the behavior of the 
SBF calibration in the near-IR \Kp\ band, where Worthey's models
predict a much weaker dependence on \v-i, but potentially larger scatter
as the effects of age and metallicity are no longer degenerate.

Stellar surface brightness fluctuations are very red, since they are
dominated by luminous red giant stars.  The advantages of
observing IR SBFs are clear:  fluctuations are ${\sim}33$ times brighter
at $K$ than at $I$, making them observable to greater distances.
The SBF amplitude is inversely proportional to the seeing FWHM, 
and the seeing is typically much better in the IR than at optical
wavelengths.
Fluctuations are also red compared to the globular cluster (GC) population,
so the contrast between stellar SBFs and GCs is higher at $K$.
Finally, stellar population models predict that $K$-band SBF magnitudes
have a much weaker dependence on \v-i\ than at $I$, reducing or eliminating
the need to accurately measure the color of the galaxy (Worthey 1993a).
  
Several studies have demonstrated the feasibility of measuring IR SBFs,
including Luppino \& Tonry (1993), 
Pahre \& Mould (1994), and JLT.
These papers report results
for a rather limited set of galaxies in the Local Group (LG) and the Virgo
Cluster, but the consistency in the measured calibration of \MK\
is encouraging  ($\MKp\,{=}\,{-}5.61\,{\pm}\,0.16$ Luppino \& Tonry,
$M_{K_{sh}}\,{=}\,{-}5.77\,{\pm}\,0.18$ Pahre \& Mould,
$\MKp\,{=}\,{-}5.62\,{\pm}\,0.29$ JLT).  
All three of these papers report absolute fluctuation
magnitudes that are consistent with Worthey's (1994) predictions
based on simple stellar population models.   
While the mean \MKp\ of JLT and Pahre \& Mould (1994) are quite consistent, 
differences in apparent fluctuation magnitudes for individual galaxies
are larger than the stated uncertainties allow.
JLT discuss several possible reasons for the disagreement in 
fluctuation magnitudes and showed that residual variance resulting
from spatial variations in the detector sensitivity 
and dark current can contribute
significantly to the power spectrum in low signal-to-noise ratio ($S/N$) 
observations.
A residual pattern
of only 0.1\% of the sky level can significantly change the fluctuation
magnitude measured.
To use IR SBFs as a reliable distance indicator, 
it is critical that the observations be of sufficiently high 
$S/N$ ratio to avoid biases from residual variances.  
JLT also demonstrated that properly subtracting
the sky and carefully sampling the point spread function (PSF) 
are crucial to 
accurately measuring the fluctuation amplitude.
An uncertainty in the background subtraction of 2\% of the sky level 
can dominate the uncertainty in the fluctuation magnitude.

Our Virgo sample showed a large dispersion (0.29 mag) due to 
the depth of the cluster and observational errors.  Pahre \& Mould
(1994) also addressed the dispersion in fluctuation magnitude
in a similarly-sized sample of Virgo ellipticals.
Clearly a larger sample of galaxies is needed to calibrate the IR SBF
distance scale.  To better understand the effects stellar populations
have on the fluctuation amplitude, we observed five galaxies in the 
Fornax cluster.  Fornax is ideal for several reasons:
first, it is much more centrally concentrated than the Virgo
cluster.
The reduced dispersion in distances allows us to better quantify
and understand the uncertainty in our SBF measurements. 
Second, Fornax contains a large number of giant elliptical galaxies
with a wide range of \v-i\ colors, metallicities, and globular cluster 
populations.  The Virgo galaxies observed by JLT
and Pahre \& Mould (1994)
spanned a very limited range in \Mg2\ index to avoid stellar population
variations which may affect the calibration of \MK. 
We now extend the sample to calibrate \Kp\ fluctuation magnitudes across
a wider range of metallicities and \v-i\ colors to compare with theoretical
stellar population models.
Finally, Cepheid distances to several Virgo cluster galaxies and to the Fornax
spiral NGC~1365 have been measured using the Hubble Space Telescope (HST). 
We can empirically anchor our \Kp-band SBF calibration both to the Cepheid 
and the $I$-band SBF distance scales.

This paper reports results from our observations of four
Eridanus cluster elliptical galaxies in addition to the Fornax galaxies.  
Eridanus is not as compact a cluster
as Fornax, but we can still compare our IR SBF results with the $I$-band
results from Tonry (1997).  
Finally, we return to our SBF measurements for the seven
Virgo ellipticals discussed by 
JLT and apply the improved techniques
we describe in this paper.  We present the updated results for these
galaxies in Section~\ref{virgorevisited}, along with new observations 
of NGC~4365.

\section{Observations and Data Reduction}

\subsection{Observations}

This paper presents results derived from data obtained during several
observing runs using the QUIRC camera 
mounted at the f/10 focus of the University of Hawaii 2.24~m telescope.
QUIRC is a near-IR camera employing the first-\ and
second-generation 
$1024^2$-pixel HgCdTe ``HAWAII'' detectors (Hodapp et al. 1996).
These observations are compared to previous work which used the 256$^2$-pixel
NICMOS detector 
(JLT, Luppino \& Tonry 1993).
The second-generation science-grade HAWAII array has far fewer bad pixels, 
higher quantum efficiency, and a much more spatially uniform response
than the earlier HgCdTe devices.
The dark current for these detectors is less than 1~e$^-$s$^{-1}$ and
the readnoise is $\sim$15~e$^-$.
The QUIRC camera has a plate scale of 0\farcs1886~pixel$^{-1}$, 
providing a field of view 3\farcm2 on a side.
The University of Hawaii \Kp\ filter used for these observations
is centered at a slightly shorter wavelength (2.11 \micron) than the
standard $K$ filter to reduce the thermal background (Wainscoat \& Cowie 1992).
The mean sky background measured using the \Kp\ filter during this run was
13.4 mag~arcsec$^{-2}$.
These observations are directly comparable to those of 
JLT and
Luppino \& Tonry (1993), which used the same \Kp\ filter.

\begin{deluxetable}{cccccc}
\tablecaption{SBF Observing Runs\label{fornaxobsdata}}
\tablewidth{0pc}
\tablehead{
\colhead{Observation}&
\colhead{Camera/} &
\colhead{Scale} & 
\colhead{Field of view}&
\colhead{Gain} &
\colhead{Galaxies/clusters} \nl
\colhead{dates} &
\colhead{detector} &
\colhead{(\arcsec)} &
\colhead{(\arcmin)} &
\colhead{(e$^-$/ADU)} &
\colhead{}}
\startdata
1993 Apr 7--11 & NICMOS & 0.375 & 1.6 & 30 & Virgo \nl
1995 May 17--20 & QUIRC-1 & 0.19 & 3.2 & 6 & NGC~4365, 4489 \nl
1995 Oct 14--17 & QUIRC-1 & 0.189 & 3.2 & 6 & Fornax, Eridanus \nl
1996 Mar 1, Apr 27--28 & QUIRC-2 & 0.189 & 3.2 & 1.85 & NGC~4365 \nl
\enddata
\end{deluxetable}

We observed five giant elliptical galaxies in the Fornax
cluster and four in the Eridanus cluster over a period of four nights in 
1995 October using the first-generation 1024$^2$-pixel array.
In this paper we reanalyze the Virgo cluster data originally presented
by JLT.
Most of the Virgo observations were made using the smaller $256^2$-pixel
NICMOS3 array (Hodapp, Rayner, \& Irwin 1992).
Two of the JLT galaxies were observed using the first-generation
1024$^2$-pixel array in the QUIRC camera. 
NGC~4365 was observed again in 1996 March and April
using the second-generation $1024^2$-pixel HgCdTe array with 
improved electronics, and we compare these newer data to those presented
by JLT.
Details of the individual observing runs are presented in 
Table~\ref{fornaxobsdata}.

\subsection{Preliminary Image Analysis and Reduction\label{reduction}}

\begin{deluxetable}{cccccccccc}
\tablecaption{Photometric Data\label{fornaxphotdata}}
\tablewidth{0pc}
\tablehead{
\colhead{Galaxy}&
\colhead{$m_1$\tablenotemark{a}} &
\colhead{$\sec z$}&
\colhead{$A_{atm}$} &
\colhead{Seeing} &
\colhead{Sky} &
\colhead{Resid. sky} &
\colhead{$t_{exp}$} &
\colhead{$t_{tot}$} &
\colhead{$A_B$\tablenotemark{b}} \nl
\colhead{NGC}& & & 
\colhead{(${{\rm mag} \over {\sec z}}$)} &
\colhead{(\arcsec)} &
\colhead{(mag/\sq\arcsec)} &
\colhead{(e$^-$s$^{-1}$pix$^{-1}$)} &
\colhead{(s)} &
\colhead{(s)} &
\colhead{(mag)}}
\startdata
Fornax&&&&&&&&&\nl
1339 & $23.43 \pm 0.10$ & 1.676 & 0.13 & 0.68 & 13.57 & $-0.25\pm0.10$ & 60 & 2160 & 0.00 \nl
1344 & $23.36 \pm 0.05$ & 1.635 & 0.10 & 0.74 & 13.15 & $-0.63\pm0.20$ & 60 & 2220 & 0.00 \nl
1379 & $23.38 \pm 0.10$ & 1.906 & 0.13 & 0.83 & 13.47 & $-0.27\pm0.17$ & 90 & 2070 & 0.00 \nl
1399 & $23.19 \pm 0.10$ & 1.888 & 0.13 & 0.83 & 13.21 & $-0.73\pm0.33$ & 90 & 1710 & 0.00 \nl
1404 & $23.36 \pm 0.05$ & 1.901 & 0.10 & 0.77 & 13.37 & $-0.80\pm0.15$ & 60 & 1980 & 0.00 \nl
\tablevspace{10pt}
Eridanus&&&&&&&&&\nl
1395 & $23.19 \pm 0.10$ & 1.621 & 0.13 & 0.75 & 13.05 & $-0.63\pm0.10$ & 90 & 1350 & 0.02 \nl
1400 & $23.36 \pm 0.05$ & 1.591 & 0.10 & 0.70 & 13.34 & $-0.40\pm0.20$ & 60 & 1980 & 0.13 \nl
1407 & $23.38 \pm 0.10$ & 1.550 & 0.13 & 0.91 & 13.56 & $-0.66\pm0.10$ & 120& 2280 & 0.16 \nl
1426 & $23.38 \pm 0.10$ & 1.865 & 0.13 & 0.89 & 13.61 & $-0.53\pm0.10$ & 90 & 1710 & 0.02 \nl
\tablevspace{10pt}
Virgo W&&&&&&&&&\nl
4365 & $22.91 \pm 0.03$ & 1.000 & 0.00 & 0.59 & 13.30 & $-0.73\pm0.21$ & 89 & 2670 & 0.00 \nl
\enddata
\tablenotetext{a}{Magnitude of a source yielding 1 \eps.}
\tablenotetext{b}{Burstein \& Heiles 1984}
\end{deluxetable}

The techniques we used to do the initial data reduction are described
by JLT.
We adopted several key improvements for the new data described
in this paper.  Flat field images and bad pixel masks were
created as described by JLT.
To improve the noise characteristics of the background for our SBF
analysis, we constructed average (rather than median) sky frames 
from ${\sim}6$ individual
sky images which were interspersed with the galaxy observations.
Stars were removed from individual images and bad pixels masked, and
the remaining good pixels averaged to form the sky frame.
The average sky frame was subtracted from individual galaxy images
prior to flat fielding and removal of bad pixels and cosmic rays.
The cleaned and flattened galaxy images were then registered to the
nearest integer pixel and the good pixels averaged as described by JLT.
Sub-pixel registration was not used because such procedures produce
noise that is correlated from pixel to pixel and is therefore not constant
with wavenumber in the power spectrum.

To calibrate the photometry, we frequently observed the 
faint UKIRT standards (Casali \& Hawarden 1992).  
The majority of our 1995 October observations were made when 
conditions were not photometric.  The photometric zero point 
fluctuated by as much as 0.25 mag during periods of all but one
of the nights.  To better calibrate our photometry, we compared
our results to photoelectric aperture photometry 
(Persson, Frogel, \& Aaronson 1979, Frogel et al. 1978, Persson et al. 1980).  
We measured the difference between our magnitudes and the photoelectric 
magnitudes for all galaxies in common (7 of 9) and used the 
mean difference and standard deviation to adjust our photometric 
zero-point and uncertainty for each night.  
Atmospheric extinction coefficients were determined by combining standard
star observations covering a range of air masses during photometric periods
of different nights. 
No color term to the photometry was measured.
Photometric zero-points and extinction coefficients are listed
in Table~\ref{fornaxphotdata}.

Next, we corrected for galactic extinction.  From Cohen et al. (1981)
we learn that
\begin{equation}
A_V : A_K : E(B-V) = 3.04 : 0.274 : 1.00. 
\label{exteq}
\end{equation}
Following Tonry et al. (1997), we take $A_V/E(B{-}V)\,{=}\,3.04$ 
for early type galaxies.
We assume that the extinction at $K$ and \Kp\ are the same, so 
combining the above relations gives $A_{\Kp}\,{=}\,0.068 A_B$.
Values for $A_B$ are listed in Table~\ref{fornaxphotdata}
(Burstein \& Heiles 1984).

No K corrections were needed for the low-redshift galaxies in this study.
In Section~\ref{wortheymodels} we describe stellar population
models used to predict fluctuation magnitudes (Worthey 1994, 1997).
The Worthey models were also used to determine the redshift dependence
of the $K$-band fluctuation magnitudes.  At a redshift of $z\,{=}\,0.005$
appropriate for the Virgo, Fornax, and Eridanus clusters, the 
K correction to the $K$-band fluctuation magnitude is $-0.001$ mag for a 17 Gyr
stellar population and +0.004 for an 5 Gyr population ([Fe/H]\,=\,0).
While these corrections are for the $K$ filter, the offsets are so
small that it is safe to assume the corrections to the \Kp\ SBF magnitudes
are negligible.
Even at redshifts of $z\,{=}\,0.025$ (7500 \kms), Worthey's models predict
corrections no larger than $\pm0.015$ mag between ages of 5 and 17 Gyr.
The fact that $K$ SBFs change so little as a function of redshift
is very encouraging.  The purpose of calibrating \Kp\ SBFs is to allow
SBF measurements to be made to much larger distances than 
possible at $I$ band, and the absence of significant K corrections 
reduces the uncertainty in doing so.

\section{Surface Brightness Fluctuation Analysis \label{howtosbf}}

\subsection{Measuring SBF Magnitudes}

We followed the same procedures described by JLT
to measure SBF amplitudes in this data, 
with important modifications to account for the contributions
from undetected globular clusters and background galaxies.  
We used the VISTA image analysis software (written by T. Lauer and
R. Stover) and additional programs
written by J. Tonry to perform the SBF analysis.
The first step in the process
was to determine the residual sky background remaining after stacking
individual images.  
We assumed that this small adjustment 
(typically ${\lesssim}0.2\%$ of the sky level) was constant
over the entire frame and could be applied as an offset to the final
flattened and registered image.
As JLT found, 
however, the fluctuation power
is quite sensitive to residual patterns in sky subtraction,  
flat fielding, and non-zero background offsets, even at the level
of ${\sim}0.1\%$ of the sky level.
We estimated the residual
sky background levels in the final images by fitting a deVaucouleurs $r^{1/4}$
profile to each galaxy using elliptical apertures.
The larger QUIRC field of view made it possible to determine
the residual sky level much more accurately than using NICMOS.
The sky levels in the JLT data were also remeasured for consistency.

The next step in measuring the SBF magnitude was to fit and subtract the galaxy
profile.  
We used a routine which allows the center, ellipticity, and position angle
to vary as we fit annuli of increasing radius, and we iterated the
procedure to ensure that the fit was well-behaved.  
Point sources were masked before fitting the galaxy profile.
The resulting model of the galaxy
was subtracted from the data.
Large-scale residual background variations were also fitted and subtracted, 
leaving a smooth, flat background with stellar SBFs, globular clusters, 
background galaxies, foreground stars, and noise superimposed.
The noise varies with position in the image because of the subtraction
of the galaxy, although in our IR images sky-subtraction noise dominates
everywhere in the image except very near the center of the galaxy. 
Globular clusters and background galaxies 
were masked as described in the following section.
The residual image was then masked with a window function which selected the
region of the galaxy to be analyzed. 
The Fourier transform and power spectrum were then 
computed for the data.
The variance arising from stellar SBFs and unresolved point sources 
was determined from the best fit of the data power spectrum with the function
\begin{equation}
P(k)~=~P_0~E(k)~+~P_1,
\label{sbffit}
\end{equation}
a linear combination of
the fluctuation power $P_0$ in \epspp\
times the expectation power spectrum $E(k)$ and 
a white noise component $P_1$ arising from photon shot noise. 
The white noise component depends on the background sky level and the mean
galaxy brightness:
\begin{equation}
P_1~=~1~+~{sky \over {\langle galaxy \rangle}}
\end{equation}
In the absence of sky background and blurring by the atmosphere,
$P_1\,{=}\,1$ and $P_0$ is simply the power at $k\,{=}\,0$. 
In real observations, the fluctuations are convolved with the PSF.
We must also account for the contribution to the power spectrum caused by the
sharp edges of our window function and point source mask.
  
The expectation power spectrum $E(k)$ is the power spectrum of the PSF
modified slightly to include the power spectrum of the window function,
scaled by the square root of the mean galaxy brightness within the window. 
To compute \E, we started by extracting a bright PSF star from the 
galaxy-subtracted image.  The region around the star was cleaned of
other stars, if necessary, and we confirmed that the background was zero.
We used as large an aperture around the star as possible without 
allowing sky subtraction noise or uncertainties in the residual background
level to corrupt the power spectrum of the
PSF at low wavenumbers.  This usually required masking the PSF for radii
greater than ${\sim}5$\arcsec.  
We checked the aperture photometry to confirm that the total flux of the PSF
star was consistent with the power measured within the aperture.
The Fourier transform and power spectrum of the PSF were computed and
fitted with a third-order polynomial.  
For small wavenumbers ($k\,{\lesssim}\,15$), the fit to the power spectrum
was used to create a composite PSF spectrum, 
and the flux at $k\,{=}\,0$ is normalized to unity. 
Next, we multiplied the window function mask by the point source mask
which removed the detected globular clusters and galaxies brighter than
the limiting magnitude.
The window mask was then multiplied by the square root of the galaxy
profile model. 
The scaling of the variance by the mean galaxy brightness was thereby
included in the expectation power spectrum. 
The Fourier transform and power spectrum for the resulting image were
computed.  The expectation power spectrum \E\ is the convolution of
the normalized power spectrum of the PSF and the power spectrum of the
modified window function: 
\begin{equation}
E(k) = |PSF(k)|^2 \otimes |W(k)|^2.
\end{equation}
In computing \E\ in this way, we are effectively multiplying the data
by the mask prior to convolution with the PSF.  
The effect of this approximation on the expectation power 
spectrum is minimal, and \E\ is very nearly the power spectrum 
of the PSF alone.

The fluctuation magnitude is computed from the fit to the data in
the radial region of the galaxy where $P_0$ is mostly 
constant with wavenumber $k$.
Near the center of the galaxy, the sampled 
area is small and the power spectrum
varies due to the small number of fluctuations being measured.
In the largest apertures, noise from sky subtraction dominates 
and the $S/N$ ratio is significantly reduced.  
To get the best fit for $P_0$, we examine the power spectra in apertures
centered on the galaxy, and include the largest area possible in the final
fit that does not compromise the quality of the power spectrum.
For the data described in this paper, we measured $P_0$ in apertures
with inner and outer radii of 2, 6, 12, 24, 48 and 80\arcsec.
The fits to the power spectra were performed in radial regions where
$P_0$ was nearly constant and for wavenumbers 
in the range $15\,{<}\,k\,{<}\,50$.
Once $P_0$ is known from the fit to the data (Equation~\ref{sbffit} above), 
The fluctuation magnitude is defined as
\begin{equation}
\mKp = {-}2.5\,\log(P_0)~+~m_1~-~A_{atm}\sec z~-~A_{\Kp},
\end{equation}
where $P_0$ is a flux in \epspp\
and $m_1$ is the magnitude of a source yielding
one \eps\ at the top of the atmosphere.

\subsection{Removing Residual Variances From Globular Clusters and Galaxies}

\subsubsection{In Theory\label{intheory}}
Pixel to pixel variance can arise from a number of sources other than
the surface brightness fluctuations resulting from the Poisson 
statistics of the stellar population. 
Foreground stars, globular clusters, other galaxies 
(usually in the background) and clumpy dust 
can all contribute to the fluctuation power we measure.  
By working in the near-infrared, we minimize the effects of dust
absorption.
Faint galactic foreground stars beyond limiting magnitude of our 
observations are not numerous enough to add significantly to the SBF
power measured (Wainscoat et al. 1992).
To remove the contribution from GCs, we
first identify as many as possible.
GCs brighter than the completeness limit are masked.
Assuming we know the form of the globular cluster luminosity function (GCLF),
we can extrapolate beyond the completeness limit 
to determine the residual variance due to undetected clusters.  
This technique has been shown to work successfully 
by decreasing the cutoff magnitude and confirming that the fluctuation
magnitude does not change significantly (Tonry 1991).   
Blakeslee \& Tonry (1995) used SBFs to measure
the GCLF of Coma cluster elliptical galaxies 
without detecting the majority of the GCs individually.  
The procedure for removing background galaxies is similar.  
We determine the magnitudes of the galaxies brighter than the completeness
limit and assume a power-law distribution of galaxies to estimate
the contribution from undetected galaxies.

\begin{figure}
\epsscale{0.8}
\plotone{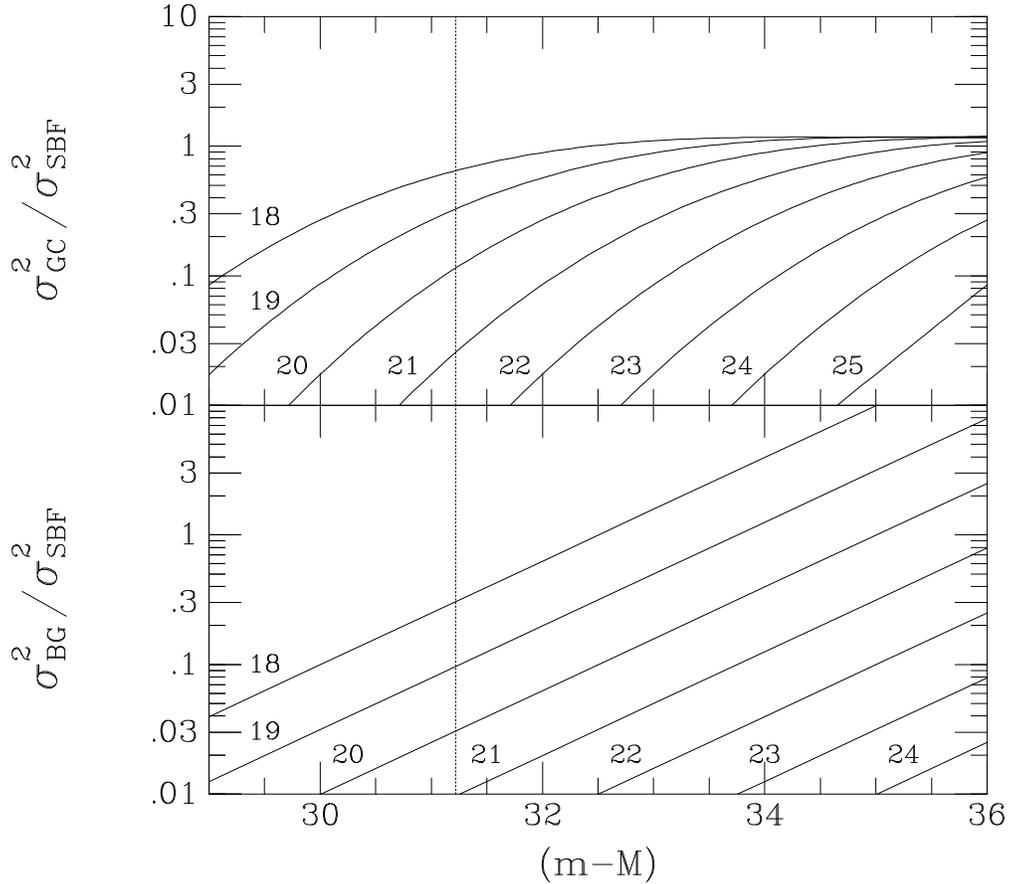}
\vspace{20pt}
\caption[Predicted Fluctuations from Globular Clusters and Galaxies]
{Theoretical models can be used to predict the relative contributions
of stellar SBFs, globular clusters, and background galaxies.  
This plot shows the spatial power (variance) resulting
from globular clusters (top panel) and background galaxies (lower panel)
as a fraction of the stellar SBF power and as a function of the limiting
magnitude of the observation.  At the distance of the Fornax cluster
(marked with a vertical line), the model predicts a contribution to
the total power spectrum from GCs that is approximately 30\% of the stellar 
SBF signal when the limiting magnitude is $\Kp\,{=}\,19$.  The ratio
of variances is 0.3 at the intersection of the vertical line and the
$\Kp\,{=}\,19$ line.  
The corresponding contribution from background galaxies would be 10\% at
the same distance and limiting magnitude.  
The following parameter values were used to construct the GC models:  
the GC specific frequency $S_N\,{=}\,5$, the galaxy color 
$(V{-}K)\,{=}\,3.3$,
and the Gaussian GCLF $\sigma\,{=}\,1.4$ mag (Blakeslee et al. 1997).
The models assume the distance modulus to Virgo of 
$(m{-}M)\,{=}\,31.0$ (Tonry et al. 1997, Ferrarese et al. 1996),
the peak of the GCLF in the $V$ band of 
$m^0_V\,{=}\,21.56$ (Blakeslee \& Tonry 1995),
and the mean fluctuation magnitude $\mKp\,{=}\,25.25$ 
at the distance of Virgo (JLT).
To construct the galaxy models we assumed a power law slope of 0.3
and a normalization of 1\ galaxy\,mag$^{-1}\,$arcsec$^{-2}$ at 
$K\,{=}\,29.38$
(Cowie et al. 1994, Djorgovski et al. 1995).  The mean 
galaxy surface brightness at \Kp\ was taken to be 17.78 mag\,arcsec$^{-2}$
and the absolute fluctuation magnitude $\MKp\,{=}\,{-}5.6$ mag (JLT).
The vertical line marks the distance modulus of the Fornax cluster
$(m{-}M)\,{=}\,31.22$. 
\label{models}}
\end{figure}

Since stellar SBFs are much redder than GCs,
we expect the relative contribution to the fluctuations 
from GCs to be smaller.  
However, we do not sample the GCLF as deeply at \Kp\ because the
background is brighter and our sensitivity to faint point sources is reduced.  
To estimate the relative contributions to the fluctuation power
from GCs, background galaxies, and stellar SBFs as a function
of limiting magnitude and distance,
we adapted Blakeslee \& Tonry's (1995) models to the $K$ band.
To calculate the models plotted in Figure~\ref{models},
we derived the following expressions for the variances
arising from stellar SBFs, GCs, and background galaxies.
In general, the variance arising from a population with $n(f)$ sources
per unit flux per pixel is
\begin{equation}
\sigma^2_{pop}~=~ \int_0^{f_{lim}}n(f)~f^2~df
\end{equation}
(Tonry \& Schneider 1988, Eq.~4).
From the definition of the fluctuation magnitude ${\overline m}$,
the stellar SBF contribution is
\begin{equation}
\sigma_{SBF}^2~=~\langle galaxy \rangle~10^{-0.4({\overline m}-m_1^*)}
\label{SBF}
\end{equation}
where $m_1^*$ is the magnitude of an object yielding one unit of flux
per total integration time.  
From Blakeslee \& Tonry (1995), Equation~12, we write the GC contribution as
\begin{equation}
\sigma_{GC}^2~=~{S_N \over 2}~\langle galaxy \rangle~
10^{0.4[\,m_1^*\,{-}\,2m^0\,{+}\,(m{-}M)\,{-}\,(V{-}K)\,{+}\,
0.8\sigma^2\ln(10)\,{-}\,15]}~{\rm erfc}(Z)
\label{GC}
\end{equation}
where $\sigma$ is the GCLF width, 
$m^0$ is the turnover magnitude for the GCLF, 
and $m_c$ is the cutoff magnitude.
All GCs brighter than the cutoff magnitude are masked and do not contribute
to $\sigma^2_{GC}$.
The argument of the complementary error function is
\begin{equation}
Z~=~{{m_c-m^0+0.8\sigma^2\ln(10)} \over {\sqrt{2}~\sigma}}.
\end{equation}
Taking the ratio of Equations~\ref{SBF} and \ref{GC} gives
\begin{equation}
{\sigma^2_{GC} \over \sigma^2_{SBF}}~=~
{S_N \over 2}\,10^{0.4[\,{\overline m}\,{-}\,2m^0\,{+}\,(m{-}M)\,{-}\,
(V{-}K)\,{+}\,0.8\sigma^2\ln(10)\,{-}\,15]}~{\rm erfc}(Z)
\label{GC/SBF}
\end{equation}
In the upper panel of Figure~\ref{models} we plotted Equation~\ref{GC/SBF} for 
several values of the 
cutoff magnitude
$m_c$ and taking the width of the GCLF to be 1.4 mag 
(Blakeslee \& Tonry 1995, 1996).
The distance scale was tied to the Virgo cluster, for which we used
a distance modulus of 31.0 mag (Tonry 1997).  
The peak magnitude for the GCLF was taken to be $23.78\pm0.16$ 
for the $V$ band at the distance of Virgo (Secker \& Harris 1993), 
and we assumed $(V{-}K)\,{=}\,2.23$ for GCs (Frogel et al. 1978) 
to get $m^0_K\,{=}\,21.55$.
We adopted a GC specific frequency of $S_N\,{=}\,5$, a typical value for giant
ellipticals (Blakeslee 1997; Blakeslee, Tonry, \& Metzger 1997; 
Blakeslee \& Tonry 1995).  
The galaxy color we adopted was $(V{-}K)\,{=}\,3.3$ 
(Frogel et al. 1978; Glass 1984).

To calculate the relative contribution to the fluctuation power from
unresolved background galaxies, we first assume a power-law distribution
\begin{equation}
n(m) = A~10^{\gamma(K-19)}.
\end{equation}
Cowie et al. (1994) found $A\,{=}\,10^4$ galaxies degree$^{-2}$mag$^{-1}$
at $K\,{=}\,19$ 
and a power-law coefficient of $\gamma\,{=}\,0.30$.  This agrees
nicely with the work of Djorgovski et al. (1995), who found 
$\gamma\,{=}\,0.315$
and the same normalization at $K\,{=}\,19$.  
We adopt $\gamma\,{=}\,0.3$, which 
translates to one galaxy~arcsec$^{-2}$mag$^{-1}$ at $K\,{=}\,29.38$.
We can write Equation~9 from Blakeslee \& Tonry (1995) as
\begin{equation}
\sigma^2_{BG}~=~{{p^2} \over {(0.8-\gamma)\ln(10)}}
10^{0.8(m_1^*-m_c)~{-}~\gamma(29.38-m_c)}.
\label{BG}
\end{equation}
The pixel scale $p$ is included so that $\sigma^2_{BG}$ is in units of
flux per integration time per pixel$^2$.
Combining Equations~\ref{SBF} and \ref{BG}, 
we find that the relative contribution from background galaxies is
\begin{equation}
{{\sigma^2_{BG}} \over {\sigma^2_{SBF}}}~=~{{p^2} \over {(0.8-\gamma)\ln(10)}}
10^{0.4[(m{-}M)~{-}~\MK~{+}~m_1^*]~{+}~(\gamma-0.8)m_c~{-}~29.38\gamma}
\end{equation}
where $(m{-}M)\,{=}\,(\overline m{-}\overline M)$ and 
$\MK\,{\approx}\,{-}5.6$ 
(JLT).
To plot realistic models as a function of cutoff magnitude $m_c$ and
distance modulus, we need to estimate typical mean galaxy brightnesses
and the magnitude of an object giving one unit of flux per integration.
Over the areas of the Fornax galaxies in our sample in which we perform
the SBF analysis, the \Kp\ surface brightness ranges from 
17.0 to 18.8 mag~arcsec$^{-2}$, which corresponds to a typical flux of 
${\sim}6$~\epspp.  For the 1995 October observing run, 
$m_1\,{\approx}\,23.35$ for a source
with a flux of 1~\eps.  The resulting plots for the Fornax observations
are shown in the lower panel of Figure~\ref{models}.

The models shown in Figure~\ref{models} are useful for predicting
the relative contributions of GCs and galaxies to our SBF measurement
for a given exposure time.  
For example, at the distance of Fornax 
(marked with a line in Figure~\ref{models}) and assuming
a cutoff magnitude of $K\,{=}\,19$, 
the contribution from unexcised
globular clusters is one-third that of the stellar SBFs, or 25\% of $P_0$.
At the same limiting magnitude and distance, the contribution from background
galaxies is smaller than that of the stellar SBFs by a factor of 10.
It is interesting to note that there is a maximum GC contribution
to the fluctuation power measured which is approximately equal to the
stellar SBF variance.  In the $I$ band, the GC contribution is a factor of
10 greater compared to the stellar fluctuations.  
Even though the effect of GCs on $P_0$ is greatly reduced at $K$, it is
not negligible.  When the contribution from GCs is larger than
${\sim}30\%$ of the fluctuation power $P_0$ 
(or the ratio in Eq.~\ref{GC/SBF} exceeds 0.5), 
the uncertainty in the correction becomes the dominant source of error and
the distance derived unreliable.  
It is therefore crucial that IR SBF measurements be deep enough to 
adequately sample the GC and background galaxy populations.
Inspection of Figure~\ref{models} shows that the ability to remove the
GC and galaxy populations from the SBF measurement will be
a limiting factor in extending the SBF distance scale to ${\sim}$100 Mpc.
In the following sections, we describe the practical techniques we used
to measure and subtract the GC and galaxy fluctuations, and argue that
identifying objects in deep optical images and removing them from IR
images prior to measuring the SBF amplitude is potentially the most
profitable way to proceed.

\subsubsection{In Practice \label{likelier}}

To measure the globular cluster and background galaxy contributions 
to the SBF power spectrum, we first identified and performed photometry on
as many
sources as possible in the image using the DoPHOT program version 3
(Schechter, Mateo, \& Saha 1993).  
DoPHOT identifies sources in a galaxy-subtracted image and determines
their magnitudes by scaling the PSF.  
For optical SBF work, a modified version of DoPHOT (ver. 2) was developed 
which carefully accounts for the galaxy subtraction and SBF contributions
to the noise (Tonry et al. 1990).  
In the results presented here, we used
the standard ``off-the-shelf'' DoPHOT version 3 
which does not correct for galaxy subtraction in modeling the noise.
At \Kp, sky subtraction noise dominates at all radii except very near
the center of the galaxy.  DoPHOT performs well except within a few
arcseconds of the center, and we examined the objects found near the center
and edited them manually as appropriate.

\begin{figure}
\plotfiddle{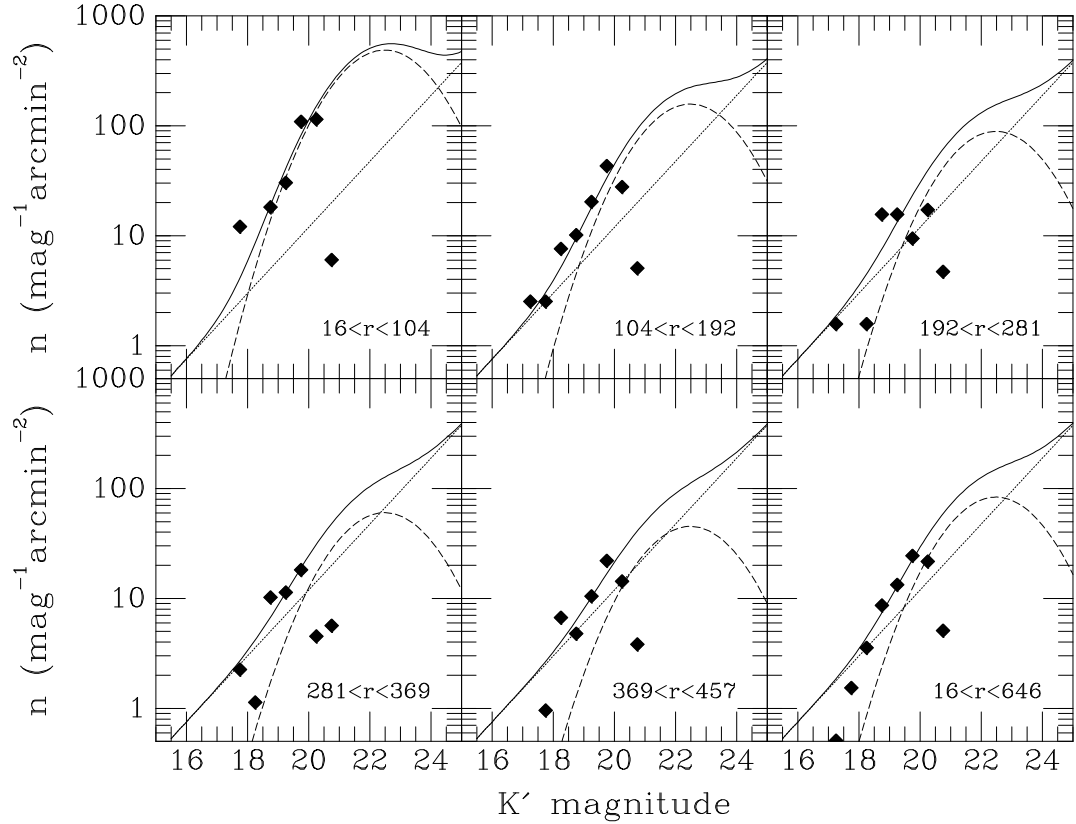}{6.0in}{270}{60}{60}{-210}{380}
\caption[Fits to the Luminosity Functions of GCs and Galaxies for NGC~4365]
{Luminosity functions of globular clusters and galaxies 
identified in the new \Kp\ QUIRC image of NGC~4365. 
Each panel shows the number density of objects as a function of 
magnitude for a separate radial region (radii are noted in pixels).
The final panel shows the number density for the entire image.  
Model fits to the globular cluster and galaxy luminosity functions
are shown as dashed lines, and their sum as a solid line.
The model was integrated to compute the residual variance from
GCs and galaxies fainter than the cutoff magnitude, near $\Kp\,{=}\,19.75$
in this case.  These fits are typical of the QUIRC observations 
with a 3\farcm2 field of view; smaller field NICMOS observations find
many fewer sources and therefore the fits to the luminosity functions
are poorly constrained.
\label{likefig}}
\end{figure}
  
With the output list of objects and magnitudes from DoPHOT, we 
constructed luminosity functions in radial bins and determined the 
completeness limit.  
A mask was made to remove all objects brighter than the completeness limit,
and we estimated the residual variance for the population fainter than
the cutoff magnitude.  To do this, we fitted the observed luminosity function
with a sum of a Gaussian GCLF and a power-law function for the background
galaxies.  
Objects brighter than ${\sim}(m^0_K{-}4)$ were not included in the
fits, nor were objects with $S/N\,{<}\,4.5$.  
The GCLF width was assumed to be 1.4 mag 
(Blakeslee \& Tonry 1995, 1996), and the distance was varied to 
determine the best fit.  When the GCLF was not well-sampled, the distance
was fixed and the fits performed without distance as a free parameter.
The galaxies were assumed to have a power-law distribution 
with an exponential coefficient of 0.30 
(Cowie et al. 1994, Djorgovski et al. 1995).  The normalization was 
left as a free parameter, and we confirmed that the fit predicted 
reasonable number counts at faint magnitudes.
Because we tried to constrain both the GC and galaxy populations 
with relatively few points at the bright end of the distributions, 
the covariance between
the components was often significant.
For these observations, the completeness limit was 2 to 3.5 mag brighter
than the peak of the GCLF.
Our ability to sample the luminosity function was not only limited by 
the bright sky background, but also by the small field of view.
We found that the fits were best constrained by fixing the
distance to the galaxy and hence the GCLF peak $m^0_K$, either using
Cepheid or $I$-band SBF distances.  The fits using these assumed 
distances were good, but our measurements of the GCLF do not constitute
an independent measurement of the distance given our sampling of only
the brightest few clusters.  
Once the luminosity functions had been fitted, we extrapolated and 
integrated the contributions to the variance coming from sources fainter
than the cutoff magnitude.  
We then computed the residual power $P_r$ for the region of
the galaxy being examined, and we subtracted $P_r$ from $P_0$ prior to 
computing the fluctuation magnitude \mKp.  

\subsubsection{Using Optical Images to Identify GCs and Galaxies\label{Imasks}}

\begin{figure}
\epsscale{0.6}
\plotone{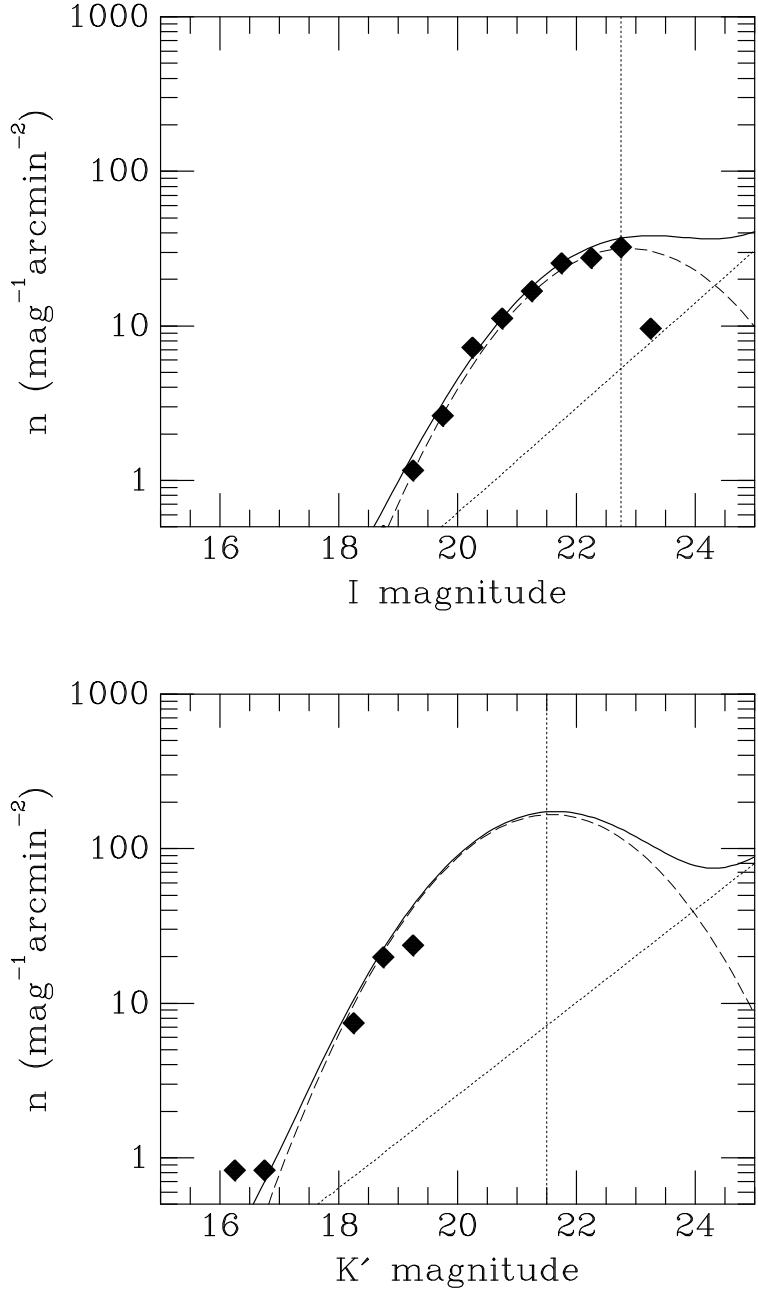}
\vspace{20pt}
\caption[Comparison of $I$ and \Kp\ Fits to the GC and Galaxy 
Luminosity Functions for NGC~1399]
{Comparison of the $I$-\ and \Kp-band fits to the globular cluster
and background galaxy luminosity functions for NGC~1399.  The $I$-band
data from Tonry (1997) provide a more secure determination of the
SBF contribution from undetected sources because it reaches a much fainter
completeness limit.  Using the $I$-band data to identify undetected sources
in the \Kp\ image reduces the GC and galaxy contribution to the \Kp\ SBF
magnitude to insignificant levels.  The completeness limit 
(marked with a vertical line) for the $I$-band
image has been translated to \Kp\ assuming $(I{-}K)\,{=}\,1.22$.
As in the previous plot, fits to the GC and galaxy luminosity functions 
are marked with dashed and dotted lines, respectively, and their sum with a 
solid line.
\label{likefig2}}
\end{figure}

IR SBF measurements are limited by the bright sky background
rather than photon shot noise in the fluctuations themselves.  
Because the background is so bright, integration times would have to
be much longer at \Kp\ than at $I$ to reach comparable limiting
magnitudes.  However, at \Kp\ fluctuations are much brighter than at $I$, 
so it is not necessary to reach as faint a limiting magnitude 
to measure the fluctuations.
The result is that at \Kp\ we identify many fewer globular clusters than
in comparable observations at $I$.  
Furthermore, modern CCDs typically cover a larger field of view than IR
detectors, 
making it possible to sample more of the GCLF than we can at \Kp.
A comparison of the luminosity functions from 
$I$ and \Kp-band SBF observations of NGC~1399 is shown
in Figure~\ref{likefig2}.  
The $I$ band data are taken from Tonry (1997).
Even though the peak of the GCLF is brighter
at \Kp\ than it is at $I$, the deeper $I$-band images sample a much greater
fraction of the GCLF.
We can take advantage of the deeper $I$-band images by using them to identify
GC and galaxy positions, and we use this information to mask sources in the 
\Kp\ image.
This allows us to remove objects that are not detectable
in the \Kp\ image before performing the SBF analysis.  
In the case of NGC~1399 (Fig.~\ref{likefig2}), 
the $I$-band completeness limit is approximately
22.75 mag.  
Given that the mean color of GCs is $(V{-}K)\,{=}\,2.23$ (Frogel et al. 1978)
and $\v-i\,{=}\,1.01$ (Ajhar \& Tonry 1994), 
we adopt a mean $(I{-}K)\,{=}\,1.22$.
The equivalent cutoff magnitude of the $I$-band observation shifted to \Kp\ 
is 21.5,
2.5 mag fainter than the completeness limit of the \Kp\ observation alone.  
Examination of Figure~\ref{models} and assuming a distance modulus of
31.22 to Fornax (Tonry 1997) shows that this corresponds to a
ratio of $\sigma^2_{GC} / \sigma^2_{SBF}\,{\lesssim}\,0.01$.  
On the other hand, had we relied upon the \Kp\ observations alone, the 
cutoff magnitude would be $\Kp\,{\approx}\,19$ mag, 
and the corresponding ratio 
$\sigma^2_{GC} / \sigma^2_{SBF}\,{\approx}\,0.3$.
When GCs and galaxies make up such a large fraction of the SBF magnitude,
the uncertainty in the $P_r$ correction is large.
By combining $I$ and \Kp\ data we gain the important advantage of 
reducing the $P_r$ correction to a negligible level.

This paper reports SBF results from our \Kp\ data alone, 
applying the corrections 
determined by fitting the GC and galaxy luminosity functions as 
described above.
We also present \Kp\ SBF magnitudes determined using the $I$-band masks.
In practice, we collected $I$-band images and object masks from the
optical SBF survey (Tonry 1997).  The $I$-band images and masks were scaled, 
rotated, and cropped appropriately to bring them into registration with
the \Kp\ image.  In the subsequent SBF analysis, we replaced the \Kp-band
mask with the $I$-band mask, removing clusters and galaxies from 
the \Kp\ image which were not detected because of the sky background
noise.  The residual variance $P_r$ was taken to be zero and the 
fluctuation magnitude computed.

\subsection{Other Sources of Residual Variance\label{residvar}}

The NICMOS3 and first-generation HAWAII 1024$^2$-pixel HgCdTe 
detectors used for these 
observations have banded sensitivity and dark-current patterns 
that are easily seen in the raw images 
(Hodapp et al. 1996, Fig.~4).  
JLT found that imperfect sky and galaxy subtraction can leave a 
residual variance which contributes to the spatial power spectrum.
In low-$S/N$ ratio measurements, the residual background variance
can be a significant fraction of the fluctuation power measured.
Imperfectly fitting the galaxy profile and background 
leave patterns that add power to the spectrum.
The \Kp\ sky background is ${\sim}100$ times brighter than the galaxy, 
even near the center,
and imperfect sky subtraction affects the fit to the galaxy profile.
Also, if an object is not masked prior to fitting the galaxy
profile, the iterative elliptical profile fitting routine will find too
much flux and overestimate the galaxy brightness at
that radius, resulting in an elliptical ring in the galaxy profile.
Unmasked objects brighter than the completeness limit 
that escaped detection by
DoPHOT are not accounted for in the fit to the 
luminosity function and also contribute to the residual power spectrum.
In fitting the SBF power spectrum to determine $P_0$, we avoid low wavenumbers
$k{\lesssim}15$ which are contaminated by large-scale features in the image.
However, residual patterns are not smoothly varying, and are observed to 
contribute power on scales of $20\,{<}\,k\,{<}\,50$, 
where the fit to $P_0$ is
commonly performed. 

We attempted to measure the contribution from residual patterns in several 
different ways.
First, we examined the radial behavior of $P_0$.  
We frequently found that $P_0$ increased with radius by as much as a 
factor of two within 1\arcmin\ of the center.
Stellar population models would require a decrease in age of 
${\sim}10$ Gyr or a increase in [Fe/H] of at least 1 dex
to account for a gradient this large (Worthey 1994, 1997).
Galaxies that do show radial SBF gradients such as 
NGC~3379 (Sodemann \& Thomsen 1995) or
Maffei 1 (Luppino \& Tonry 1993) only show a change of ${\sim}0.2$ mag 
within an arcminute of the galaxy center.
Furthermore, from dynamical arguments, we expect the outer envelopes of
elliptical galaxies to be older and more metal-poor than the centers
(Faber et al. 1995).
Stellar population models predict that IR SBFs should
be fainter in the outer regions, not brighter, if the outer regions of
the galaxy are older and metal deficient.
We regard the increasing SBF amplitude with radius as evidence 
for residual variance, especially in relatively low-$S/N$ ratio 
observations.   
Variance remaining due to improper galaxy subtraction, incomplete 
point-source masking,
and residual dark current patterns will affect the larger outer annuli 
more strongly, leading to an increasing fluctuation power with radius.
Assuming a modest residual power of 0.1\% of the sky value usually
brings all apertures to the same fluctuation power.
 
We conducted several experiments to see if 
a level of residual variance of 0.1\% of the sky level was reasonable.
In one trial,
we took two averaged sky images observed on one night,
separated in time by about half an hour.
The resulting difference image contains residual patterns at the level
we might expect to accumulate during typical observing, 
since in our normal image reduction strategy we 
averaged individual sky frames taken during approximately half an hour. 
The power spectrum of the difference image was measured using the 
same SBF procedures used to analyze the data images.
The spectrum rose steeply for $k\,{<}\,20$, and gradually decreased
with increasing $k$.
The residual variance measured in this way should be considered an 
upper limit, since the patterns were added coherently  
(sky frames are combined without registration).  
In the final registered galaxy images, residual patterns were partially
cancelled because individual galaxy images had been dithered.
In another experiment, we combined individual sky images 
with their relative offsets, so that residual patterns were not 
added coherently as in the previous example.  
We then computed the power spectrum and measured the
deviation from a flat spectrum.
The power spectra show the same behavior observed in the previous experiment
but with a lower amplitude.
Since this measurement of the residual variance is for
the sky subtraction only, and does not account for any residual pattern 
from the galaxy subtraction or point source removal, 
it is a lower limit of the residual variance we
expect in a galaxy power spectrum.
The variances measured in these two experiments fall between 
0.02\% and 0.2\% of the sky value, confirming that 0.1\% is a
reasonable estimate of the residual spatial variance.

Occasionally faint objects were not identified by DoPHOT; 
we measured the flux in these objects and 
found that they can contribute up to 0.1\% of the sky level in variance. 
We removed such objects and repeated the SBF analysis to confirm the
reliability of our procedure for removing residual variance.  
A few faint unidentified objects appear when using 
the $I$-band point source mask
because very red galaxies detected at \Kp\ can be fainter than the 
completeness limit in the $I$-band image, and therefore remain unmasked.  
Our measurements of residual power in the sky images 
were used as guidelines for determining the best value of the 
residual variance ($P_g$), 
which we determined by eliminating any radial trend in $P_0$.

When radial gradients in $P_0$ could be attributed to residual variance,
adopting a correction of ${\sim}0.1\%$ of the sky value usually removed the
effect.  However, residual patterns were not always uniform, 
and low-$S/N$ ratio observations frequently showed variable radial behavior 
that could not be removed
by simply subtracting a uniform residual variance.  Experience with our
data showed that the best way to deal with
the residual variance was to use the value of $P_g$ that removed a slope in
$P_0$ when a clear trend exists, and to
adopt a uniform correction of 0.1\% of the 
sky level prior to sky subtraction when the radial behavior of $P_0$ was 
uncertain.  
Minimizing the radial variation in $P_0$ usually required a correction
of $P_g\,{\approx}\,0.2\,P_0$ at $\p0/p1\,{=}\,3$, or in general, 
$P_g\,{\approx}\,0.5\,P_1$.
The lower the background noise, the smaller the required correction for
residual variance. 
The SBF analysis was performed in the annular region that produced the
cleanest power spectrum, and the residual variance correction $P_g$
was subtracted from the fluctuation power $P_0$ prior to computing the
fluctuation magnitude (in addition to any correction $P_r$ 
applied to account for faint globular clusters and galaxies).
Since the value of $P_g$ is only an estimate
of the residual spatial variance, its associated uncertainty is
large when $P_g$ is a significant fraction of $P_0$.

Clumpy dust distributions can also add to the fluctuation power measured
and cause the distance to the galaxy to be underestimated.
By working at \Kp, we minimize the effects of dust extinction. 
Careful inspection of the \Kp\ images revealed no evidence of patchy
dust, and we made no additional correction to \mKp.
If clumpy dust were present in significant quantities, we would expect
it to be concentrated toward the center of the galaxy.
The resulting SBF amplitude would decrease with radius, which is opposite
the trend arising from the other sources of residual variance 
described in the previous paragraph.
For our data, we find that SBF amplitudes are larger in the outer annuli,
and we conclude that patchy dust in the innermost regions of these
galaxies does not affect the \Kp\ SBF magnitude.

\section{SBF Results}

\subsection{Fornax and Eridanus\label{fornaxanderidanus}}

\begin{figure}
\epsscale{0.9}
\plotone{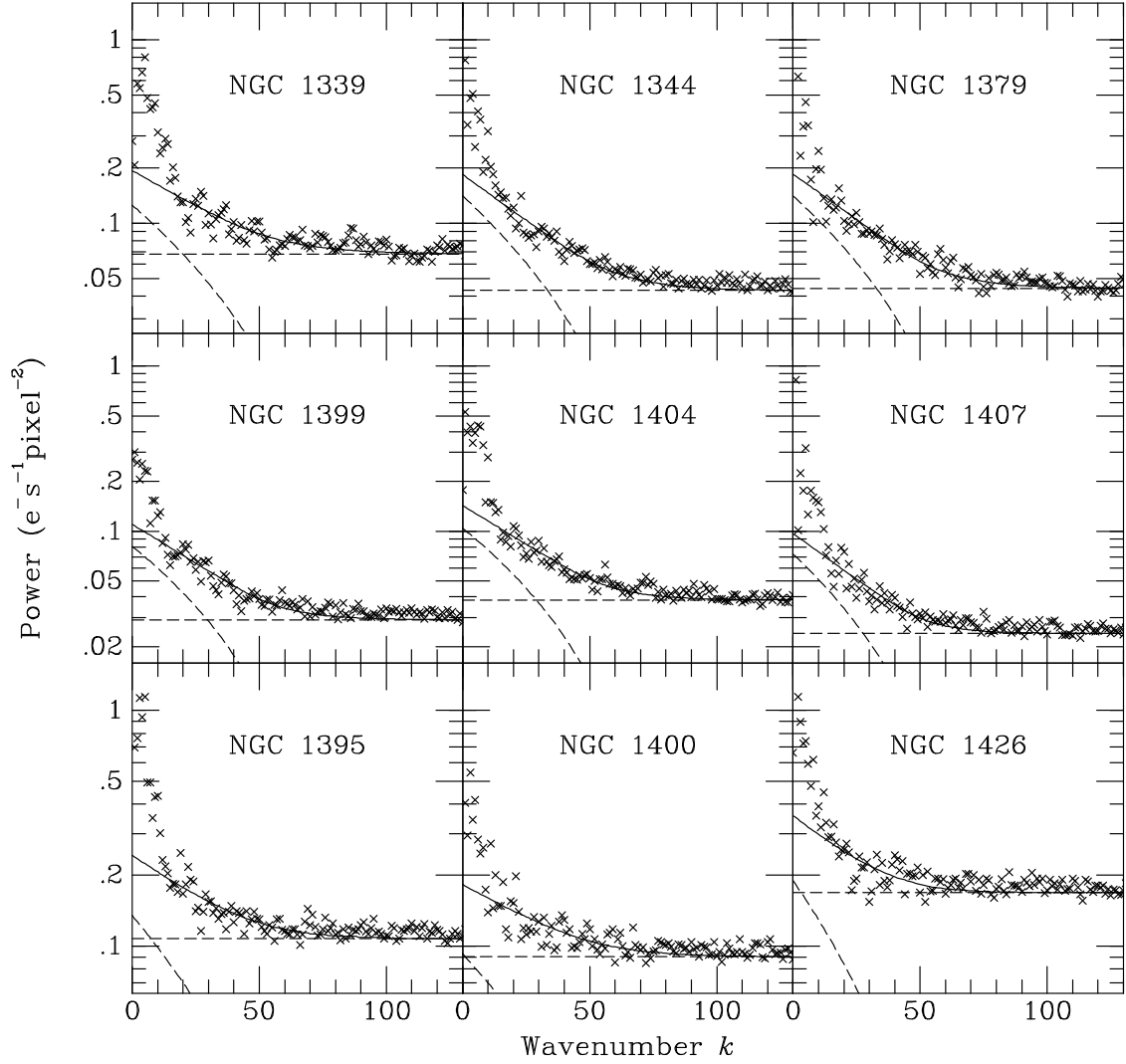}
\vspace{20pt}
\caption[SBF Power Spectra for Fornax and Eridanus Cluster Galaxies]
{Power spectra for the Fornax and Eridanus cluster 
galaxies.
The dashed lines are the individual components $P_0{\times}E(k)$ and $P_1$,
and the solid line indicates their sum.  Powers are in units of \epspp.
The Fourier transforms were computed in 512$^2$-pixel sub-images. 
\label{powerspectra}}
\end{figure}

\begin{figure}
\plotone{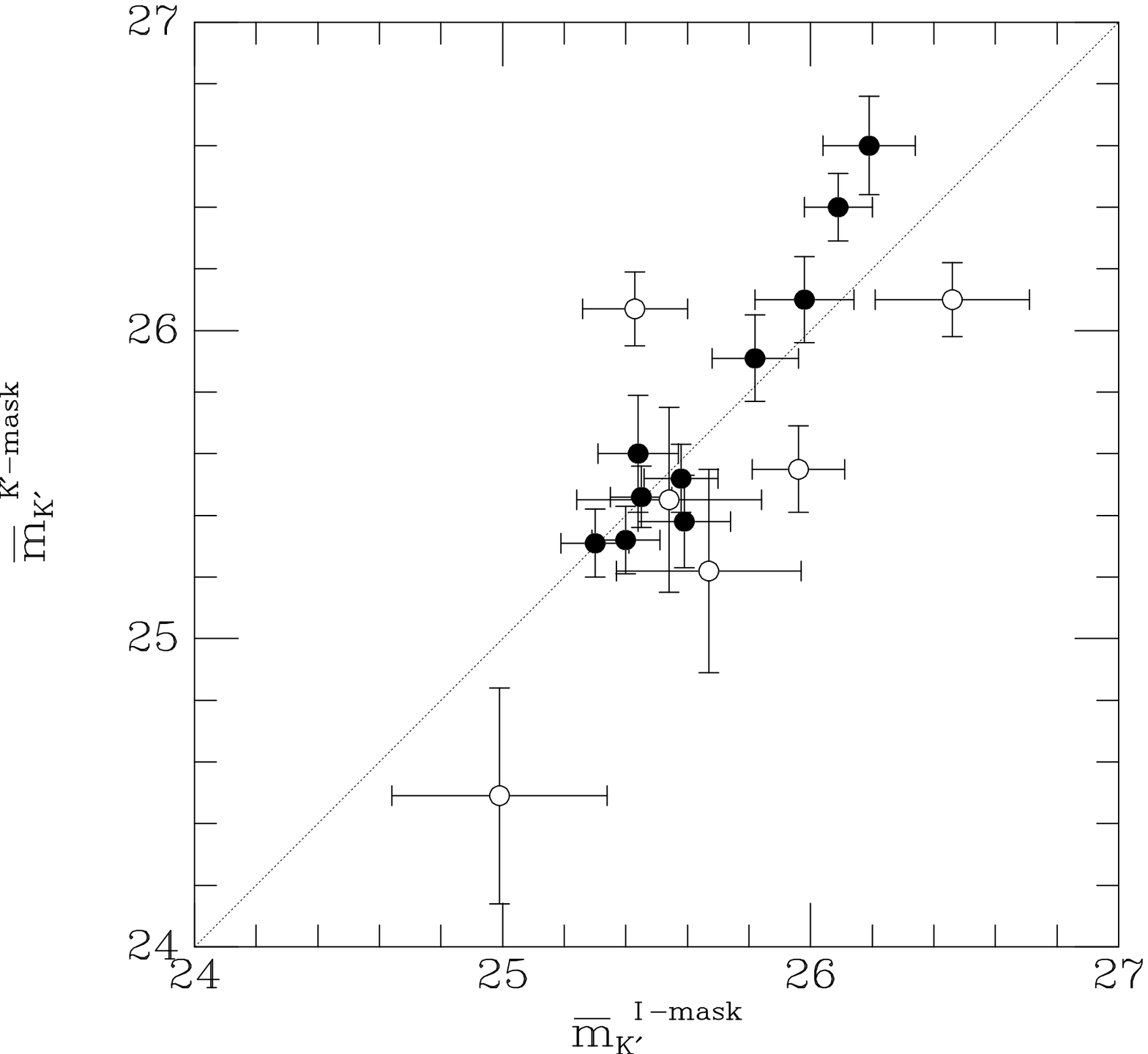}
\vspace{20pt}
\caption[Comparison of \Kp\ SBF Magnitudes Using IR and Optical GC Masks]
{Comparison of fluctuation magnitudes \mKp\ using $I$ and \Kp\ globular
cluster and galaxy masks.  The dotted line passes through the origin and
has a slope of 1, and is not a fit to the data.  Low-$S/N$ ratio points
are plotted with open symbols.
The error bars on the 
\Kp-mask data are incomplete because no residual variance corrections 
were made for these measurements.  
\label{imaskvskmask}}
\end{figure}

We measured apparent fluctuation magnitudes \mKp\ for five giant elliptical
galaxies in the Fornax cluster using the techniques described in 
Section~\ref{howtosbf}.
The power spectra are shown in Figure~\ref{powerspectra}, along with the
individual components of the fit $P_0{\times}E(k)$ and $P_1$.
Two values of \mKp\ are listed in Tables~\ref{fornaxmbarskp} and 
\ref{fornaxmbarsi} for each galaxy:
the magnitudes in Table~\ref{fornaxmbarskp} were derived using the
\Kp\ data alone;
the magnitudes in Table~\ref{fornaxmbarsi} were determined using the 
$I$-band images to identify and remove faint GCs and galaxies.
We compared the fluctuation magnitudes measured using the two different
masks in Figure~\ref{imaskvskmask}.  

In the first set of \mKp\ magnitudes (Table~\ref{fornaxmbarskp}), 
we fitted the background galaxy and
GC luminosity functions and integrated them to compute the
power $P_r$ from objects fainter than the completeness limit.
Because of the high background, the \Kp\ observations only sample the
brightest GCs and galaxies (Fig.~\ref{likefig2}), 
and $P_r$ was frequently a significant fraction of the total power $P_0$.  
When $P_r/P_0$ exceeds ${\sim}0.3$, the uncertainty
in the SBF magnitude from the GC and galaxy contribution becomes the 
dominant source of error.  
We did not apply an additional 
residual variance correction $P_g$ because of the 
significant uncertainty in $P_r$, thus the uncertainties listed in
Table~\ref{fornaxmbarskp} are not directly comparable to those in 
Table~\ref{fornaxmbarsi}.

For the second set of fluctuation magnitudes (Table~\ref{fornaxmbarsi}), 
the deeper $I$-band images were used to identify and mask GCs and galaxies,
and we assumed that their contribution to \mKp\ is negligible ($P_r\,{=}\,0$).
As was shown in Section~\ref{Imasks} for NGC~1399, the cutoff
magnitude in the $I$-band images (shifted to \Kp) is usually 2 mag fainter
than the typical cutoff magnitude of $\Kp\,{\approx}\,19.5$ 
in the \Kp\ images.  
Applying the models in Section~\ref{intheory} (Fig.~\ref{models}), we
find that $P_r/P_0\,{\approx}\,0.01$ and can safely be ignored.
Without worrying about GCs and galaxies, we addressed the issue of 
residual variance in the \Kp\ images.  
We followed the procedures outlined in Section~\ref{residvar} and adopted
a residual variance correction $P_g$ at a level 
which minimized radial variations in $P_0$, usually near 0.1\% 
of the sky brightness.  
The relative sizes of the corrections ($P_g/P_0$) are listed in 
Table~\ref{fornaxmbarsi}.  
Figure~\ref{imaskvskmask} shows that SBF magnitudes measured using the
different masks are consistent for the high-$S/N$ observations (filled
symbols).  The diagonal line with unit slope and zero intercept 
indicates perfect agreement, not a fit to the data.  
Four of the six low-$S/N$ measurements have significantly brighter
values of \mKp\ using the \Kp\ mask than using the $I$ mask.

\begin{deluxetable}{ccccccc}
\singlespace
\tablecaption{Fornax, Eridanus, and Virgo Cluster \Kp\ 
SBF Magnitudes\label{fornaxmbarskp}}
\tablewidth{0pc}
\tablehead{
\colhead{Galaxy}&
\colhead{Radius} &
\colhead{$P_0$} & 
\colhead{$P_0/P_1$} &
\colhead{$P_r/P_0$} &
\colhead{$\mKp$}&
\colhead{$\pm$} \nl
&
\colhead{(\arcsec)} &
\colhead{$({\rm e}^-{\rm s}^{-1}{\rm pixel}^{-1})$} & & 
& & }
\startdata
Fornax&&&&&&\nl
NGC~1339 &  2--24 &$ 0.131 \pm 0.004 $& 4.5 & 0.12 & 25.55 & 0.14  \nl
NGC~1344 &  2--48 &$ 0.133 \pm 0.005 $& 3.5 & 0.12 & 25.52 & 0.11  \nl
NGC~1379 &  2--24 &$ 0.171 \pm 0.007 $& 2.3 & 0.40 & 25.60 & 0.19  \nl
NGC~1399 &  2--48 &$ 0.088 \pm 0.002 $& 3.2 & 0.38 & 26.10 & 0.14  \nl
NGC~1404 & 12--48 &$ 0.112 \pm 0.006 $& 3.4 & 0.29 & 25.91 & 0.14  \nl
\tablevspace{10pt}
Eridanus&&&&&&\nl
NGC~1395 & 12--48 &$ 0.159 \pm 0.009 $& 1.6 & 0.64 & 26.07 & 0.12  \nl
NGC~1400 &  2--48 &$ 0.097 \pm 0.005 $& 1.8 & 0.30 & 26.10 & 0.12  \nl
NGC~1407 & 12--48 &$ 0.089 \pm 0.004 $& 3.8 & 0.53 & 26.60 & 0.16  \nl
NGC~1426 &  2--48 &$ 0.209 \pm 0.021 $& 1.4 & 0.30 & 25.22 & 0.33  \nl
\tablevspace{10pt}
Virgo&&&&&&\nl
NGC~4365 &  2--48 &$ 0.053\pm 0.001 $& 4.9 & 0.25 & 26.40 & 0.11 \nl
NGC~4406 &  2--48 &$ 0.104\pm 0.008 $& 9.0 & 0.06 & 25.46 & 0.10 \nl
NGC~4472 &  2--48 &$ 0.115\pm 0.004 $&20.0 & 0.03 & 25.31 & 0.11 \nl
NGC~4489 &  2--48 &$ 0.271\pm 0.007 $& 3.1 & 0.03 & 24.49 & 0.35 \nl
NGC~4552 &  2--48 &$ 0.110\pm 0.004 $&11.0 & 0.03 & 25.32 & 0.11 \nl
NGC~4578 & 12--24 &$ 0.263\pm 0.020 $& 1.3 & 0.02 & 25.45 & 0.30 \nl
NGC~4636 &  2--48 &$ 0.113\pm 0.003 $& 4.1 & 0.09 & 25.38 & 0.15 \nl
\enddata
\end{deluxetable}

\begin{deluxetable}{cccccccc}
\singlespace
\tablecaption{\Kp\ SBF Magnitudes Using $I$-band GC Masks\label{fornaxmbarsi}}
\tablewidth{0pc}
\tablehead{
\colhead{Galaxy}&
\colhead{Radius} &
\colhead{$P_0$} & 
\colhead{$P_0/P_1$} &
\colhead{$P_g/P_0$} &
\colhead{\snsbf} &
\colhead{$\mKp$}&
\colhead{$\pm$} \nl
&
\colhead{(\arcsec)} &
\colhead{$({\rm e}^-{\rm s}^{-1}{\rm pixel}^{-1})$} & & 
& & & }
\startdata
Fornax&&&&&&&\nl
NGC~1339 & 12--24 &$ 0.119 \pm 0.005 $& 1.8  & 0.34 & 0.7 & 25.96 & 0.15  \nl
NGC~1344 &  2--48 &$ 0.141 \pm 0.005 $& 3.3  & 0.22 & 1.5 & 25.58 & 0.12  \nl
NGC~1379 &  2--24 &$ 0.140 \pm 0.004 $& 3.2  & 0.15 & 1.8 & 25.44 & 0.13  \nl
NGC~1399 &  2--48 &$ 0.088 \pm 0.003 $& 3.0  & 0.32 & 1.1 & 25.98 & 0.16  \nl
NGC~1404 & 12--48 &$ 0.105 \pm 0.004 $& 2.8  & 0.17 & 1.6 & 25.82 & 0.14  \nl
\tablevspace{10pt}
Eridanus&&&&&&&\nl
NGC~1395 & 12--48 &$ 0.134 \pm 0.007 $& 1.3  & 0.22 & 0.8 & 25.43 & 0.17  \nl
NGC~1400 &  2--48 &$ 0.094 \pm 0.004 $& 1.1  & 0.48 & 0.4 & 26.46 & 0.25  \nl
NGC~1407 &  2--48 &$ 0.072 \pm 0.004 $& 3.0  & 0.15 & 1.7 & 26.19 & 0.15  \nl
NGC~1426 &  2--48 &$ 0.195 \pm 0.016 $& 1.1  & 0.51 & 0.4 & 25.67 & 0.30  \nl
\tablevspace{10pt}
Virgo&&&&&&&\nl
NGC~4365 &  2--48 &$ 0.053 \pm 0.002 $& 4.5  & 0.00 & 4.5 & 26.09 & 0.11  \nl
NGC~4406 &  2--48 &$ 0.107 \pm 0.007 $& 8.7  & 0.06 & 5.6 & 25.45 & 0.10  \nl
NGC~4472 &  2--48 &$ 0.116 \pm 0.006 $&16.0  & 0.03 &11.3 & 25.30 & 0.11  \nl
NGC~4489 &  2--48 &$ 0.409 \pm 0.019 $& 1.7  & 0.41 & 0.6 & 24.99 & 0.35  \nl
NGC~4552 &  2--48 &$ 0.106 \pm 0.004 $& 4.9  & 0.09 & 3.0 & 25.44 & 0.11  \nl
NGC~4578 & 12--24 &$ 0.147 \pm 0.005 $& 1.4  & 0.41 & 0.5 & 25.54 & 0.30  \nl
NGC~4636 &  2--48 &$ 0.097 \pm 0.005 $& 3.5  & 0.13 & 2.0 & 25.59 & 0.15  \nl
\enddata
\end{deluxetable}

Fluctuation magnitudes were also measured for the four ellipticals in the 
Eridanus cluster, following the same procedures as those used to
measure SBFs in the Fornax galaxies (Tables~\ref{fornaxmbarskp} and 
\ref{fornaxmbarsi}).  
Because the Fornax observations were our first priority,
somewhat less time was allocated to the Eridanus galaxies.  
Eridanus is also more distant, so the resulting observations
have significantly lower in their $S/N$ ratios.  
Only NGC~1407 has a ratio of $\p0/p1\,{>}\,3$.  

Estimating uncertainties for \mKp\ can be difficult.  We proceeded by 
first identifying all the probable sources of error:  the statistical 
uncertainty in the fit for $P_0$, errors in measuring the sky offset,
errors in fitting the galaxy profile, uncertainties
in subtracting sources of residual variance, and so forth.
We modified the parameters individually through their full ranges and 
measured the change in $P_0$.  From the change in $P_0$ we estimated the
1-$\sigma$ uncertainty arising from the parameter in question.  We then
combined the uncertainties in quadrature to get the uncertainties listed
in Tables~\ref{fornaxmbarskp} and \ref{fornaxmbarsi}.  
This procedure assumes that
the different sources of uncertainty are independent, an assumption which
is not strictly true.  For example, modifying the residual sky level
affects the fit to the galaxy profile slightly.  
Errors in fitting the galaxy profile affect our estimate of residual
spatial variance as well.
In the procedure used to fit the luminosity functions of the globular
clusters and background galaxies, an error in one component obviously
affects the fit to the other.  While the cross correlations are obviously
not zero, they were found to be small compared to the principal errors.  
The uncertainties computed are reasonable when we compare them with
other distance estimates to calibrate the \Kp\ SBF distance scale.

\subsection{Virgo Revisited\label{virgorevisited}}

We reanalyzed the data originally presented by JLT using the updated 
SBF techniques presented in this paper.
We improved our \Kp\ SBF measurements by determining the globular cluster
and background galaxy contributions to the fluctuation amplitude.
We also calculated magnitudes using the masks generated from the 
$I$-band images and applied a residual variance correction,
just as we did for the Fornax and Eridanus galaxies. 
New fluctuation magnitudes are listed in Tables~\ref{fornaxmbarskp} 
and \ref{fornaxmbarsi}.
Observations of five Virgo cluster galaxies are higher in 
$S/N$ ratio than the Fornax and Eridanus data, 
and it is reassuring to find that the GC, galaxy, and residual pattern 
corrections are quite small.  The difference between fluctuation magnitudes
measured using the different masks is negligible.  
In the NICMOS data, the field of view was small enough that very few 
GCs were observed, and the SBF $S/N$ ratio was high enough that the GC 
contribution was negligible.

Three of the Virgo galaxies had low $S/N$ ratios, 
two of which appeared to have
anomalously bright fluctuations (JLT, Pahre \& Mould 1994).
In our reanalysis of the JLT 
Virgo data, NGC~4578 was found to have a \mKp\
typical for the Virgo cluster but a large uncertainty due to the low $S/N$
ratio.
The fluctuation magnitudes reported for NGC~4365 by JLT 
and Pahre \& Mould (1994) were anomalously bright, 
if this galaxy resides in the 
Virgo~W cloud behind the main Virgo cluster core 
(Tonry et al. 1990, Forbes 1996).
We re-observed this galaxy in 1996 
using the science-grade HAWAII detector
and applied the full corrections for
the globular cluster and background galaxy contributions. 
The new observations 
confirmed that the earlier measurement
was biased by a significant GC population masquerading as stellar SBFs.  
Applying the GC correction to the 1995 data (JLT)
brought it
into agreement with the 1996 observation
reported here.
Applying the $I$-band GC mask allowed us to further check the 
consistency of the \Kp\ measurement.  
Magnitudes listed in Tables~\ref{fornaxmbarskp} and \ref{fornaxmbarsi} 
are for the new 1996 data.
When we used the $I$-band mask to remove GCs from the
\Kp\ image, we measured a fluctuation magnitude of 
$\mKp\,{=}\,26.09\,{\pm}\,0.11$ mag.  
NGC~4365 illustrates that SBF amplitudes can be too large,
making the galaxy appear significantly closer than it really is,
if exposure times are not long enough to achieve a sufficiently high
$S/N$ ratio and the globular cluster contribution is not adequately measured.
With no {\it a priori} knowledge of the
distance to NGC~4365 and follow-up observations, the 
JLT data 
would have lead to the incorrect conclusion that this galaxy resides in the 
main Virgo cluster.

JLT and Pahre \& Mould (1994) also observed an anomalously bright fluctuation 
magnitude in NGC~4489.  We have not yet re-observed this galaxy, and 
reanalysis of the JLT data
does not expose an obvious GC contribution
that could be responsible for the fluctuation magnitude measured.

\subsection{The Effect of Low ${\bf S/N}$ Ratios on IR SBF Measurements}

\begin{figure}
\plotone{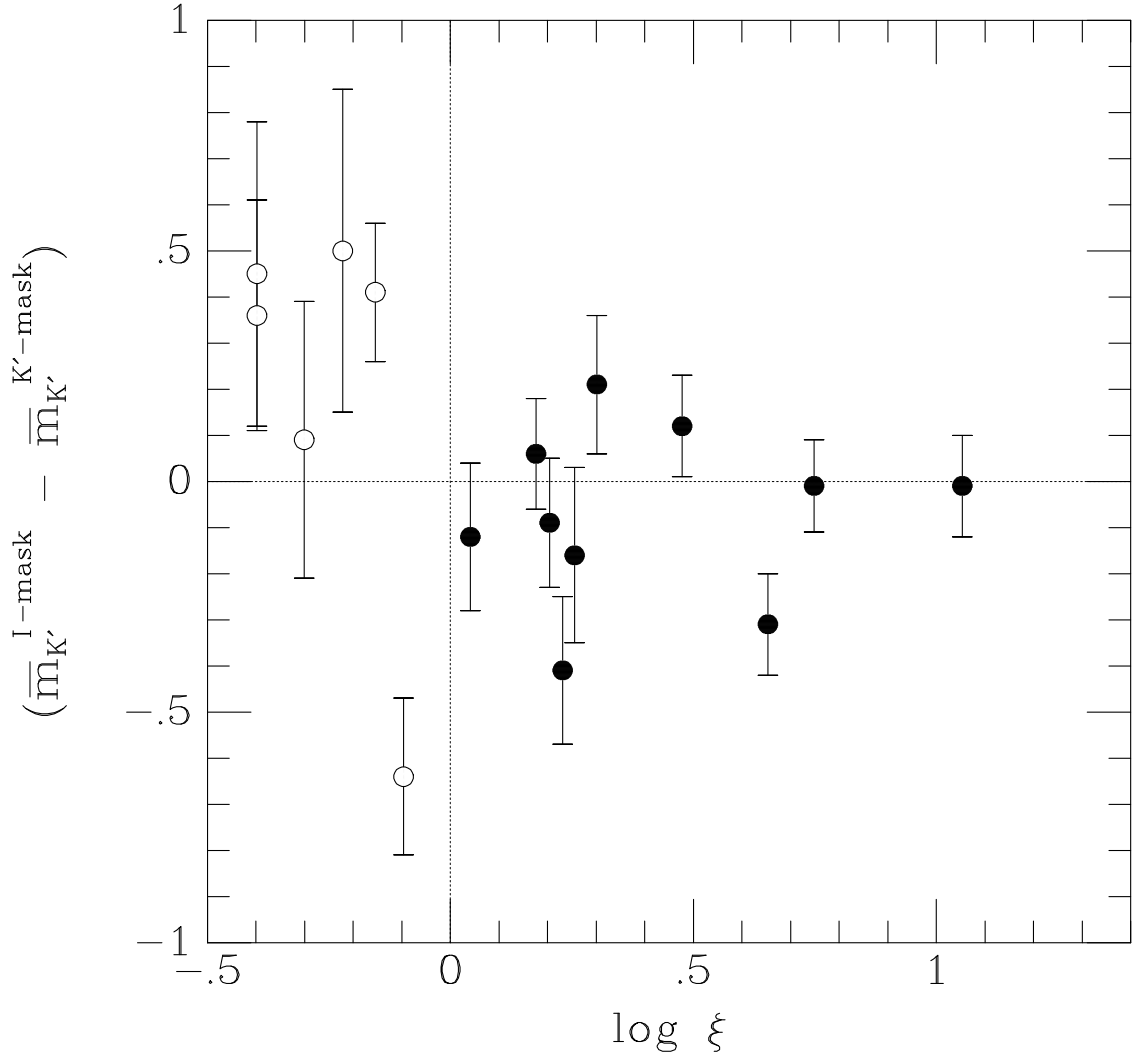}
\vspace{20pt}
\caption[Difference Between Fluctuation Magnitudes as a Function of $S/N$]
{The differences between fluctuation magnitudes \mKp\ measured using 
the $I$ and \Kp\ globular cluster masks are plotted as a function of 
log(\snsbf).  We plot the logarithm of the $S/N$ ratio because \mKp\ is
proportional to log($P_0$) and \snsbf\ is proportional to $P_0$.
Low-$S/N$ ratio measurements show the greatest differences between
measurements using different masks.  Measurements with 
$\snsbf\,{<}\,1$ (open
symbols to the left of the vertical line)
were not included in the calibration of \MKp. \label{snfig}}
\end{figure}

We have referred to \p0/p1 as the $S/N$ ratio.  With a better understanding
of the variance arising from sources other than the stellar SBFs, we 
computed a more realistic measure of the $S/N$ ratio:
\begin{equation}
\snsbf = {{(P_0-P_r-P_g)}\over{(P_1+P_g)}}
\label{snsbfeq}
\end{equation}
Values for \snsbf\ are listed in Table~\ref{fornaxmbarsi}. 
Galaxies for which $\snsbf\,{<}\,1$ were not included
in the calibration of \MKp.  Comparison of the fluctuation magnitudes
listed in Tables~\ref{fornaxmbarskp} and \ref{fornaxmbarsi} showed that the SBF
magnitudes for these low-$S/N$ ratio galaxies are unreliable, in agreement
with the conclusion reached 
by JLT 
from analysis of the Virgo data alone.
The differences are plotted as a function of $\log(\snsbf)$ 
in Figure~\ref{snfig}.
The mean difference between \mKp\ values for the low-$S/N$ ratio 
observations is 0.41 mag;  for the galaxies with $\snsbf\,{>}\,1$ 
the difference
is only 0.15 mag, comparable to the individual uncertainties.  
The highest-$S/N$ ratio observations show the smallest
difference between values of \mKp\ measured with and without the $I$-band
point-source masks.
It is clear that when either $P_r$ or $P_g$ exceeds ${\sim}0.3P_0$,
the resulting \mKp\ is unreliable.  In the following section,
only the observations with $\snsbf\,{>}\,1$ (or $\p0/p1\,{\gtrsim}\,3$)
were included in the statistical analysis.  
From our observations
we conclude that   
it is crucial that future IR SBF observations be deep enough to ensure
that all sources of variance can be removed and that the $S/N$ ratio of
the {\it stellar} SBFs is larger than 1.

The definition of \snsbf\ in Equation~\ref{snsbfeq} is empirical, but we
can also use the results of our survey to predict \snsbf\ as a function 
of observational parameters.  The relationship can be approximated as
\begin{equation}
\snsbf \approx 0.4\,D\,\theta^{-1/2}\,\left({{1000~\kms} \over {cz}}\right)
\left({{t_{exp}} \over {1000~{\rm s}}}\right)^{1/2}
\label{predictedsnsbf}
\end{equation}
where $D$ is the telescope diameter in meters 
and $\theta$ is the seeing FWHM in arcseconds.
The relationship between \snsbf\ and the observational parameters is not
particularly good because a number of other parameters also affect the 
measured value of \snsbf.  Equation~\ref{predictedsnsbf} provides
a basic guideline for the minimum integration required to make $K$-band
SBF measurements.  The coefficient in Equation~\ref{predictedsnsbf} is
somewhat less than the value of 0.5 
found by Pahre \& Mould (1994),
and we emphasize that achieving a $\snsbf\,{=}\,1$ is marginal and assumes
that residual variances have been subtracted.
To make reliable $K$-band SBF distance measurements, we 
recommend that on-source integration times be
\begin{equation}
t_{exp} > 6\,D^{-2}\,\theta\,\left({{cz} \over {1000~\kms}}\right)^2\,
\times 1000~{\rm s}.
\end{equation}

\section{The \Kp\ SBF Distance Scale Calibration\label{calibration}}

\subsection{Adopted Distances\label{distances}}

To calibrate the \Kp\ SBF distance scale, we first calculated absolute
fluctuation magnitudes \MKp\ for the high-$S/N$ ratio galaxies in our sample.
We adopted three different (but not independent) sets of distances: 
Cepheid distances to M31 and the Virgo cluster, 
$I$-band SBF distances to individual galaxies,
and cluster $I$-band SBF distances averaged  
over a larger sample of galaxies than we observed at \Kp.  

Our calibration of the IR SBF distance scale 
is strengthened significantly by the Cepheid distances to several Virgo
spirals that have now been measured (Freedman et al. 1994; Pierce et al. 1994;
Ferrarese et al. 1996; Saha et al. 1996a,b; Sandage et al. 1996).  
Of the five reported Cepheid distance measurements, four were made using data 
acquired using the Hubble Space Telescope.  
Three of these HST measurements are remarkably consistent:  
NGC~4321 $(m{-}M)\,{=}\,31.04\,{\pm}\,0.17$ (Ferrarese et al. 1996), 
NGC~4496a $(m{-}M)\,{=}\,31.03\,{\pm}\,0.14$ (Saha et al. 1996b), 
NGC~4536 $(m{-}M)\,{=}\,31.10\,{\pm}\,0.13$ (Saha et al. 1996a); 
NGC~4639 is ${\sim}1$ mag more distant than the other Virgo Cepheid distances
(Sandage et al. 1996).
Like NGC~4365, it is probably a member of the background Virgo W~cloud, 
and has been excluded from the average Virgo Cepheid 
distance modulus used to calibrate the \Kp\ SBF distance scale.
We adopt a cluster distance of 
$(m{-}M)\,{=}\,31.06\,{\pm}\,0.13$ ($d\,{=}\,16.27$ Mpc) for the Virgo cluster.

Recently the HST Key Project Team has measured a Cepheid distance
to the Fornax spiral galaxy NGC~1365.
At this time the analysis is nearly complete, and the Cepheid distance
modulus to NGC~1365 is approximately $(m{-}M)\,{=}\,31.30$ 
(Silbermann 1998).
In the near future HST Cepheid distances to three Fornax cluster galaxies
will be measured by the Key Project team.  
Until then, we use only the Cepheid distances to M31 and the Virgo
cluster to calibrate the \Kp\ SBF distance scale.

J. Tonry and collaborators are in the process of publishing the results
of an extensive $I$-band SBF survey (Tonry et al. 1997).
The $I$-band SBF distance scale was calibrated using Cepheid distances to 
a number of galaxies out to 20 Mpc.
We can use $I$-band distances to calibrate the \Kp\ SBF scale 
using a larger sample of galaxies than using the \Kp\ SBF 
and Cepheid distances alone.
We used the $I$-band data in two ways.
First, distances to individual galaxies (Tonry 1997) gave absolute fluctuation 
magnitudes directly for each galaxy.
Second, mean distances to each cluster (Tonry et al. 1997) were used,
allowing us to take advantage of a larger sample but requiring us to assume
that all the galaxies in a cluster are at the same distance.
In both cases the $I$-band distances were calculated according to the 
calibration of Tonry et al. (1997).
$I$-band distance moduli are listed in Table~\ref{fornaxm-Mtab}.

Recent measurements by the Hipparcos astrometric satellite 
have suggested that a systematic error in
the Cepheid distance scale may exist because of an error in the parallax
distances to the nearest stars (Feast \& Catchpole 1997).  
If this is the case, all Cepheid
distances need to be increased by ${\sim}10\%$, although the exact change
may depend on metallicity corrections to the period-luminosity relation
for Cepheid variables (Madore \& Freedman 1997).  
The Cepheid zero-point correction also affects the
$I$-band SBF distances, since they are also calibrated using the Cepheid
distance scale.  Because this issue has not yet been resolved, we adopt the
distances cited above with a caution to the reader that the \Kp\ SBF distance
calibration presented in this 
paper
may need to be revised in the future.

\subsection{Theoretical Stellar Populations\label{wortheymodels}}

Surface brightness fluctuations can also be calibrated using theoretical models
by computing the second moment of the luminosity function for a stellar
population and normalizing by the first moment.  
If we know the absolute magnitudes of the brightest stars 
(which dominate the second moment of the luminosity function) 
from stellar population models, we can determine the distance directly
from the apparent SBF magnitude.  
Alternatively, we can check our empirical calibration of the SBF distance
scale against the synthesized stellar population models.  
G. Worthey produced a set of single-age, single-metallicity 
stellar population models and computed fluctuation magnitudes in the 
Johnson $J$, $H$, and $K$ bands (Worthey 1994, 1993b).
He kindly created new models specifically for the \Kp\ filter,
and we compare our fluctuation magnitudes with these new models (Worthey 1997).
The \Kp\ SBF models show only slight differences from the $K$ models, but
are significantly different from the $K$ models transformed to \Kp\ 
using the standard relation from Wainscoat \& Cowie (1992). 
SBFs are considerably redder than the stars upon which
the Wainscoat \& Cowie transformation is based.
The new \Kp\ model magnitudes are listed in Table~\ref{kpmodels}.
The fact that the new \Kp\ models are similar to the $K$ models is
reassuring.  The detailed stellar population models Worthey created 
include the CO absorption bandhead, which falls close to the edge
of the $K$ window.
We were not certain to what extent differences in CO absorption 
between the $K$ and \Kp\ filters would affect the measured SBF magnitudes.  
Now that Worthey (1997) has produced models precisely for our filter, we can
compare them to our observations with confidence.
The relatively small difference (${\lesssim}0.03$ mag)
between $K$ and \Kp\ SBF magnitudes allows
us to compare to Pahre \& Mould (1994, $K_{\rm sh}$) as well since
$K_{\rm sh}$ is centered midway between $K$ and \Kp.
We make a detailed comparison of our data with Worthey's models in the 
following section.

\begin{deluxetable}{crrrr}
\singlespace
\tablecaption{\Kp\ Fluctuation Magnitudes from Stellar Population Models 
\label{kpmodels}}
\tablewidth{0pc}
\tablehead{
\colhead{Age (Gyr)} &
\colhead{[Fe/H]}&
\colhead{\Mg2} & 
\colhead{\v-i}&
\colhead{\MKp}}
\startdata
5&$	-0.22$ &0.185 &	1.121 &$-5.75$ \nl
&$	 0.00$ &0.215 &	1.186 &$-5.83$ \nl
&$	 0.25$ &0.268 &	1.291 &$-5.78$ \nl
&$	 0.50$ &0.316 &	1.355 &$-5.98$ \nl
8&$	-2.00$ &0.034 &	0.800 &$-3.90$ \nl
&$	-1.50$ &0.056 &	0.846 &$-4.11$ \nl
&$	-1.00$ &0.092 &	0.887 &$-4.47$ \nl
&$	-0.50$ &0.160 &	1.083 &$-5.13$ \nl
&$	-0.25$ &0.194 &	1.147 &$-5.66$ \nl
&$	 0.00$ &0.239 &	1.258 &$-5.73$ \nl
&$	 0.25$ &0.290 &	1.346 &$-5.75$ \nl
&$	 0.50$ &0.331 &	1.395 &$-5.94$ \nl
12&$	-2.00$ &0.042 &	0.857 &$-3.81$ \nl
&$	-1.50$ &0.064 &	0.892 &$-4.03$ \nl
&$	-1.00$ &0.100 &	0.930 &$-4.38$ \nl
&$	-0.50$ &0.171 &	1.110 &$-5.08$ \nl
&$	-0.25$ &0.214 &	1.204 &$-5.55$ \nl
&$	 0.00$ &0.258 &	1.304 &$-5.70$ \nl
&$	 0.25$ &0.314 &	1.396 &$-5.76$ \nl
&$	 0.50$ &0.350 &	1.427 &$-5.96$ \nl
17&$	-2.00$ &0.046 &	0.897 &$-3.78$ \nl
&$	-1.50$ &0.069 &	0.937 &$-4.15$ \nl
&$	-1.00$ &0.115 &	0.975 &$-4.47$ \nl
&$	-0.50$ &0.187 &	1.159 &$-5.08$ \nl
&$	-0.25$ &0.231 &	1.247 &$-5.47$ \nl
&$ 	 0.00$ &0.275 &	1.342 &$-5.64$ \nl
&$	 0.25$ &0.335 &	1.444 &$-5.76$ \nl
&$	 0.50$ &0.374 &	1.482 &$-5.95$ \nl
\enddata
\end{deluxetable}

In addition to single-age, single-metallicity populations, we also
considered simple composite populations.  We used Worthey's (1994)
$K$-band models to show that models vary smoothly between grid points.
We tried combining a (8, $-0.25$) (Gyr, [Fe/H]) population with a (17, $-0.5$)
population, with the younger population weighted by 5\% to 50\% (by mass).  
The point in \MK\ halfway between the two models was reached when 30\% of the 
population was younger.
We also combined a (12, $-0.25$) model with a (5, 0.0) stellar population
and found the midpoint to be at the same ratio of populations.
In other words, at least 30\% of the stellar population by mass must be
young for the young stars to dominate the $K$-band SBF magnitude.
An old population will not look like a significantly younger population in
$K$-band fluctuation magnitude if only a few percent of the mass is in the
young population. 
Stellar populations younger than 5 Gyr were not considered.
Because fluctuations are dominated by the brightest stars in the population, 
model SBF magnitudes are sensitive to the treatment of the complex 
stellar physics of the luminous asymptotic giant branch (AGB) stars.  
Fluctuations significantly brighter than model predictions may be
evidence for young populations with brighter
AGB stars than the model population.

G. Worthey's models (1997) also predict the redshift dependence of the
SBF magnitudes as a function of age.  In Table~\ref{Kcorr} we list
the K corrections in the $K$ band for redshifts between 
$0\,{\leq}\,z\,{\leq}\,0.03$.
The models are all for solar metallicity [Fe/H]\,=\,0.
The tabulated values must be added to apparent $K$-band SBF magnitudes
to find the equivalent magnitude at $z\,{=}\,0$.  The K corrections for 
$K$-band SBFs are {\it very}\ small, 
especially for the older stellar populations.
For the galaxies in this study, K corrections can be ignored. 
Even in future $K$-band SBF observations out to 10,000 \kms, uncertainties
due to K corrections will be negligible.

\begin{deluxetable}{ccccc}
\singlespace
\tablecaption{$K$-band SBF K corrections\label{Kcorr}}
\tablewidth{0pc}
\tablehead{
\colhead{$z$} &
\colhead{5 Gyr}&
\colhead{8 Gyr} &
\colhead{12 Gyr} & 
\colhead{17 Gyr}}
\startdata
0.005 & 0.004 & 0.003 & 0.002 & $-0.001$ \nl
0.010 & 0.008 & 0.006 & 0.003 & $-0.003$ \nl
0.015 & 0.011 & 0.008 & 0.003 & $-0.005$ \nl
0.020 & 0.013 & 0.010 & 0.003 & $-0.008$ \nl
0.025 & 0.015 & 0.012 & 0.003 & $-0.011$ \nl
0.030 & 0.017 & 0.013 & 0.002 & $-0.015$ \nl
\enddata
\end{deluxetable}

\subsection{Disentangling the Effects of Age and Metallicity}

\subsubsection{Fluctuation Magnitudes and $(V{-}I)$\ Color
\label{vminusisection}}

\begin{figure}
\epsscale{0.8}
\plotone{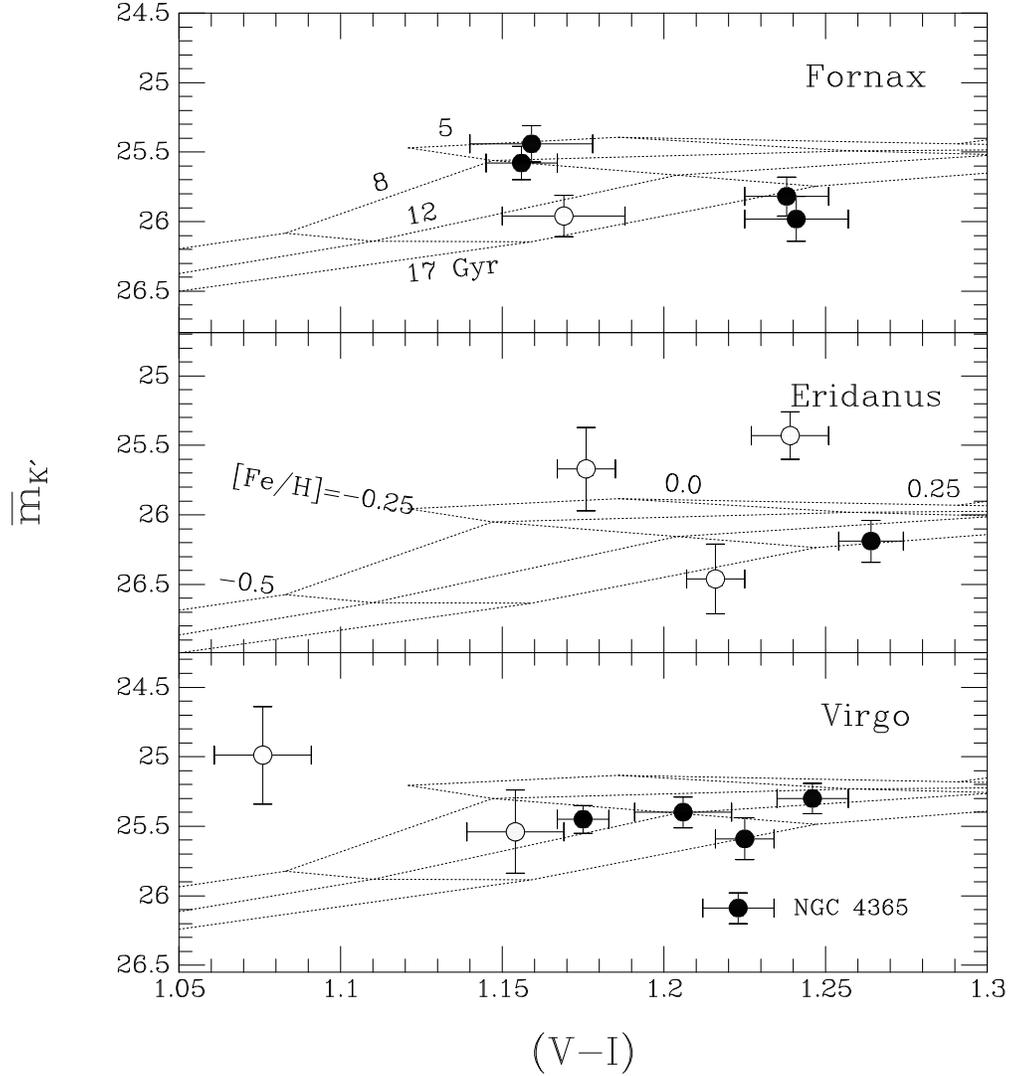}
\vspace{20pt}
\caption[Fluctuation Magnitudes as a Function of \v-i\ Color]
{Apparent fluctuation magnitudes \mKp\ are plotted as a function of \v-i\ 
color for three clusters.  
Low-$S/N$ ratio measurements are indicated by open circles.
\Kp\ stellar population models (Worthey 1997) are labeled according
to age and metallicity.  NGC~4365 resides in the W cloud, behind the main
Virgo cluster core.  The majority of the galaxies studied have fluctuation
magnitudes consistent with older stellar populations (12 to 17 Gyr), but 
those with bluer optical colors show evidence of younger stellar populations.
Low-$S/N$ ratio observations (open symbols) are unreliable.
\label{vminusi}}
\end{figure}

Broad-band colors are frequently used to constrain the properties of
stellar populations.
In the optical $I$ band, the effects of age and metallicity on SBFs 
are degenerate.
Therefore the $I$-band SBF distance scale can be calibrated 
using the \v-i\ color to parameterize both the age and metallicity of
stellar populations (Tonry 1997, Worthey 1993a).
$I$-band SBFs are sensitive to galaxy color and
\MI\ has a fairly steep slope with \v-i, but the scatter is small and $I$-band
SBFs can be used to measure accurate distances over a wide range of 
stellar population.  
At \Kp, stellar population models (Worthey 1993a, 1997) predict a shallower
slope with \v-i, but more importantly, the slope is opposite in sign.
Age and metallicity effects do not counteract each
other as in the $I$-band.  Because the age-metallicity degeneracy
is partially broken, the dispersion in fluctuation magnitudes is also
expected to be larger than at $I$.  
To explore the behavior of \Kp\ SBFs with \v-i\ color, we plotted the
fluctuation magnitudes \mKp\ as a function of \v-i\ in Figure~\ref{vminusi}.
Filled circles denote those measurements for which the $S/N$ ratio was large
($\snsbf\,{>}\,1$ or $\p0/p1{\geq}2.8$).  
The \mKp\ magnitudes derived using the $I$-band
masks were used to create the figures in this section.
We superimposed Worthey's \Kp\ models (1997), with grid lines
labeled by age and metallicity.  To place the models, we used the mean 
Cepheid distance for Virgo and cluster $I$-band SBF distances for the Fornax
and Eridanus clusters.

\begin{figure}
\epsscale{0.8}
\plotone{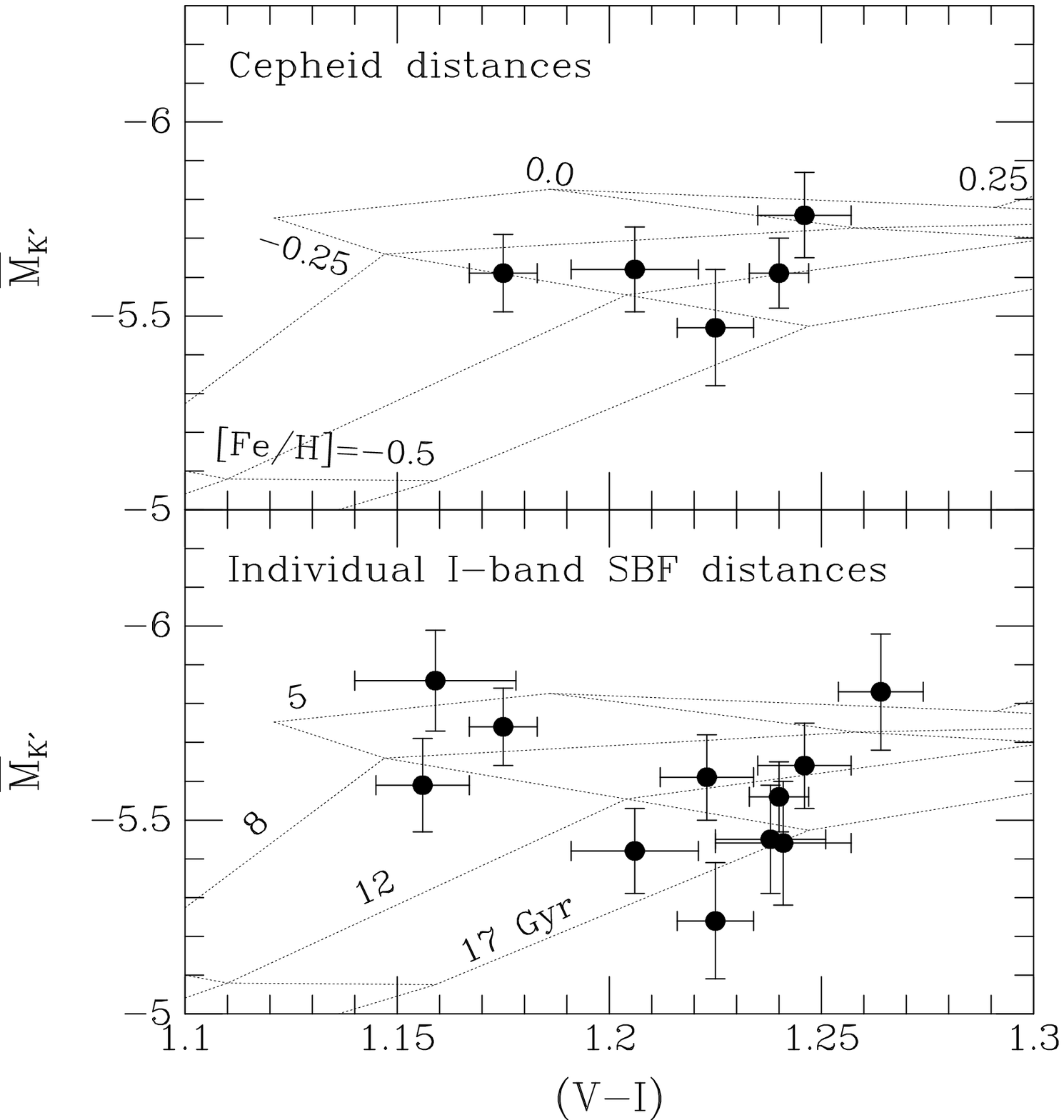}
\vspace{20pt}
\caption[Absolute \Kp\ Fluctuation Magnitudes as a Function of \v-i\ Color]
{Absolute fluctuation magnitudes \MKp\ are plotted as a 
function of \v-i\ color, 
with the Worthey (1997) models superimposed.  The top panel adopts the
Cepheid distances to M31 and the Virgo cluster, while the bottom panel
uses individual $I$-band SBF distance measurements from Tonry (1997)
for all the high-$S/N$ ratio galaxies.  Using the mean cluster
$I$-band SBF distances from Tonry et al. (1997) 
changes the lower panel only slightly.  The trend 
toward younger ages in the bluer galaxies is apparent, as is the obvious
lack of a negative slope in \MKp\ with increasing \v-i\ 
predicted by the models.   
\label{Mkvi}}
\end{figure}

From Figure~\ref{vminusi} we learn that the galaxies on the red end of
the sample are mostly consistent with older, 12 to 17 Gyr stellar populations
with metallicities near [Fe/H]\,{=}\,$-0.25$.  The bluer galaxies with 
$\v-i\,{\approx}\,1.15$ have fluctuation magnitudes consistent with 5 to 8 Gyr 
population models and similar metallicities.
The trend is easily seen in Figure~\ref{Mkvi}, in which we plotted the 
absolute fluctuation magnitudes \MKp\ as a function of \v-i\ for all the
high-$S/N$ ratio observations in our sample.  The upper panel in 
Figure~\ref{Mkvi} used the Cepheid distances to M31 and the Virgo
galaxies; the lower panel was plotted using the $I$-band SBF distances
to individual galaxies and includes galaxies in Fornax and Eridanus.
For the bluer galaxies to be consistent with 12 to 17 Gyr population models,
the SBF magnitudes would have to be at least 0.5 mag fainter.
We see that the dispersion in \v-i\ color appears to be due to 
variations in population age, not metallicity.   
This conclusion contradicts the hypothesis that the dispersion in 
colors of elliptical galaxies is the result of metallicity variations
(Buzzoni 1995).   
On the other hand, the age spread we infer from our
data is consistent with the ``bursty'' formation scenario in 
which ellipticals form episodically, as predicted by the hypothesis
that ellipticals form through mergers (Faber et al. 1995).  
Of course our sample is not very large, and this conclusion should be 
confirmed by additional high-$S/N$ ratio IR SBF measurements over a 
larger range of galaxy color.

We also learn something from the galaxies whose fluctuation magnitudes
are not consistent with the stellar population models shown.
JLT and Pahre \& Mould (1994) both noted two galaxies with anomalous 
\Kp\ SBF magnitudes in the Virgo cluster (NGC~4489 and NGC~4365).  
Our latest observations of NGC~4365 reveal that this galaxy is indeed in 
the background W~cloud. 
The new fluctuation magnitude \mKp\ is fully consistent with the 
$I$-band distance (Tonry 1997) 
and with the distance derived from the peak of the GCLF (Forbes 1996).
As stated in Section~\ref{virgorevisited} above, 
failure to remove significant residual variances (from globular clusters
in this case) can result in significantly biased fluctuation measurements.
NGC~4489's low-$S/N$ ratio fluctuation magnitude \mKp\ is 
${\sim}1$ mag brighter than the model predictions.
Either it is a young galaxy with a possibly anomalous AGB population, 
or the measurement is biased by
residual variances which have not been adequately removed. 
NGC~4489 does not appear to have an extensive GC population as does
NGC~4365, but it is unusual in other ways (JLT). 
Until deeper $K$-band images are obtained for NGC~4489,
our conclusions regarding an anomalous stellar population 
will be uncertain.  
Observations of Fornax and Eridanus cluster galaxies have revealed
other candidates with unusually bright \Kp\ SBF magnitudes; 
all have low $S/N$ ratios $\snsbf\,{<}\,1$.  
Comparison of the fluctuation magnitudes for 
NGC~1395 in Tables~\ref{fornaxmbarskp} and 
\ref{fornaxmbarsi} reveals a difference of
0.64 mag.  A reasonable correction for globular clusters from the \Kp\
observations brings the \mKp\ into agreement with the stellar
population models.  Nevertheless, the $I$-band mask should remove
the GCs much better than using the \Kp\ data alone.
NGC~1426 is also seen to be anomalously bright in both 
Tables~\ref{fornaxmbarskp} and \ref{fornaxmbarsi}.  
Given that these observations all have low $S/N$ ratios, 
all appear too bright, and all have large (and uncertain) 
corrections for residual variances $P_r$ and $P_g$, we cannot draw
any conclusions regarding their stellar populations.
When $\snsbf\,{<}\,1$, fluctuation magnitudes are unreliable
and are not included in the calibration of \MKp.

The high-$S/N$ ratio data plotted in Figure~\ref{Mkvi} can be fitted 
to find the
relationship between \MKp\ and \v-i.  The fits were performed using
the three distance estimates and the
two different sets of \mKp\ values described in 
Sections~\ref{fornaxanderidanus} and \ref{distances} above.  
We used an iterative least-squares approach to weight each point
according to the uncertainties in both \mKp\ and \v-i.
The results are listed in Table~\ref{fornaxfitstable}, which includes
the number of points used in the fit and $\chi^2$ per degree of 
freedom.
The $I$-band GC masks produce calibrations of \MKp\ 
with consistent zero points.
Whether we use the individual $I$-band SBF distances or the mean distances
for the clusters makes little difference.
Somewhat greater scatter is observed in the zero points of the 
fits which use the \Kp-mask data, confirming that the $I$-band GC and
galaxy masks produce more consistent fluctuation magnitudes.
The $\chi^2$ values indicate that our uncertainties are reasonable.
The Cepheid fits include only 5 points, and we believe it is small number
statistics, rather than overestimated errors, that causes 
$\chi^2/(n{-}2)\,{<}\,1$.
Using the $I$-band SBF distances results in larger numbers of 
galaxies in the fits.  
Differences in age and metallicity between the galaxies produce a
dispersion in \MKp\ somewhat larger than our uncertainties, and the 
$\chi^2$ per degree of freedom is larger than one.
The only statistically significant slopes we measured resulted from
using the \Kp\ GC masks and the $I$-band SBF distances.  The positive
slope suggests that \Kp\ SBFs are {\it fainter}\ in redder galaxies,
opposite the trend predicted by the stellar population models 
(Figure~\ref{Mkvi}).
No significant slope is seen in the data using the Cepheid distances alone, 
but only 5 points were included in this fit, so the lack of a correlation
is not conclusive either.
On the other hand, there is no convincing reason that a slope opposite
that predicted by the models should exist.
Our best \Kp\ SBF results are produced using the $I$-band masks,
and the fits to these data are most consistent with a constant \MKp\ across
the range in \v-i\ spanned by our sample.  
This is a very useful feature, 
since it implies that accurate \v-i\ colors do not need to
be measured to determine \Kp\ SBF distances.  
Future observations of a
significantly larger sample including 
bluer galaxies will be required to determine whether or not
\Kp\ SBFs get fainter as galaxies get bluer, 
as predicted by the stellar population models.

\begin{deluxetable}{lccccccc}
\singlespace
\tablecaption{Fits:  $\protect\MKp = a + b[\protect\v-i - 1.2]$ 
\label{fornaxfitstable}}
\tablewidth{0pc}
\tablehead{
\colhead{Distances} &
\colhead{Mask}&
\colhead{$a$} & 
\colhead{$\pm$}&
\colhead{$b$} &
\colhead{$\pm$} &
\colhead{$\chi^2/(n{-}2)$} &
\colhead{$n$}}
\startdata
Cepheids\dotfill          & $I$ &$ -5.61 $& 0.06 &$-0.9$& 1.8 & 0.8 & 5\nl
$I$ SBF individual\dotfill&     &$ -5.61 $& 0.04 &  1.4 & 1.1 & 1.9 & 11\nl
$I$ SBF group\dotfill     &     &$ -5.55 $& 0.04 &  1.7 & 1.1 & 1.2 & 9\nl
\tablevspace{10pt}
Cepheids\dotfill          & \Kp &$ -5.65 $& 0.06 &$-0.8$& 1.8 & 0.6 & 5\nl
$I$ SBF individual\dotfill&     &$ -5.56 $& 0.04 &  3.0 & 1.2 & 1.0 & 10\nl
$I$ SBF group\dotfill     &     &$ -5.53 $& 0.05 &  3.0 & 1.2 & 2.0 & 9\nl
\enddata
\end{deluxetable}
  
\begin{figure}
\plotone{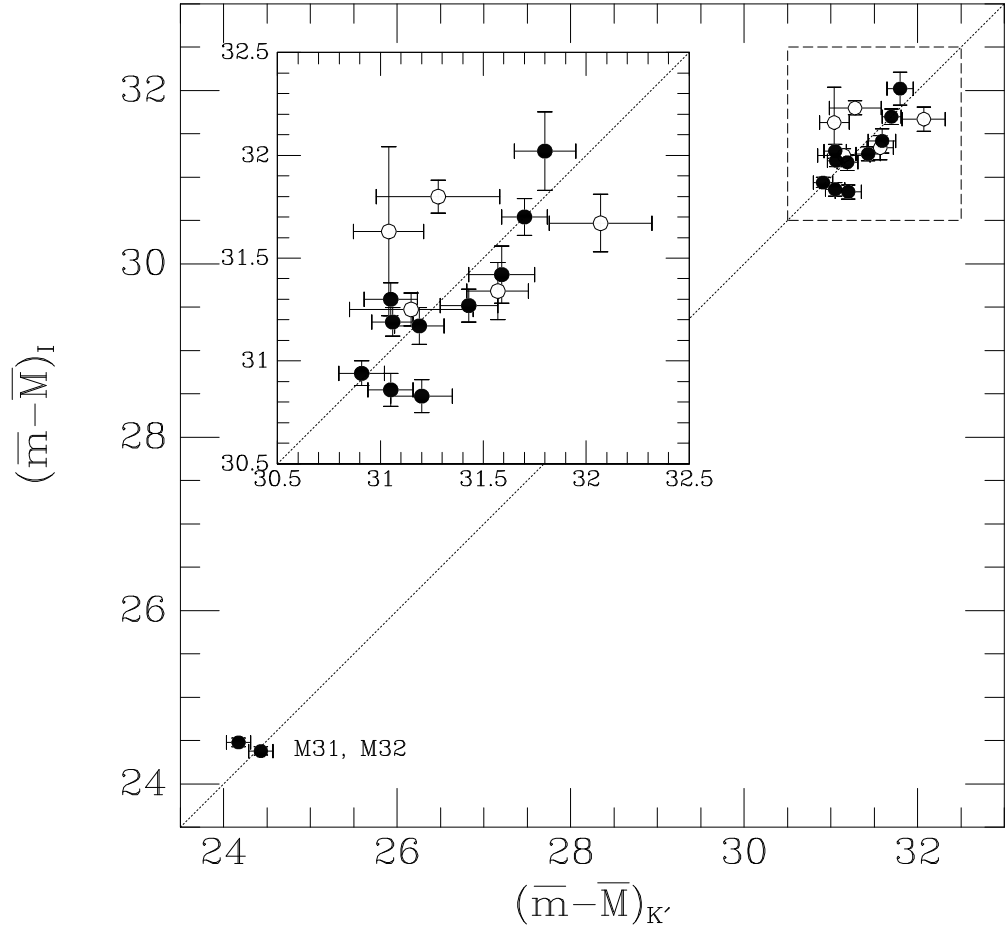}
\vspace{20pt}
\caption[$I$ and \Kp\ SBF Distance Moduli Compared]
{We plotted the $I$ and \Kp\ SBF distance moduli against each other
to demonstrate the reliability of our high-$S/N$ measurements.
The dotted line has a slope of 1 and passes through the origin;  it
is not a fit to the data.  The Virgo, Fornax, and Eridanus cluster
data are shown on an expanded scale in the inset box.
Local Group measurements are taken from Luppino \& Tonry (1993).
\label{distmodfig}}
\end{figure}

The $I$-band masks produce the most reliable fluctuation 
magnitudes, and the $I$-band SBF and Cepheid distance scale calibrations
are equally good. 
We therefore adopt as the calibration of the \Kp\ SBF distance scale
\begin{equation}
\MKp = -5.61 \pm 0.12
\label{Mkpcalibration}
\end{equation}
with no \v-i\ term for the limited range $1.15\,{<}\,\v-i\,{<}\,1.27$.  
The statistical uncertainty is 0.06 mag;
adding 0.10 mag in quadrature for the Cepheid zero point
uncertainty results in a total uncertainty of 0.12 mag.
The statistical uncertainty in the zero point was derived by fitting a 
slope to \v-i.  The scatter due to variations in stellar population
results in somewhat greater uncertainties when no correction for \v-i\ is made.
Our calibration is consistent with the JLT measurement,
which is not
surprising given that the same Virgo data were used in both studies.
This value for \MKp\ is the same as that measured by 
Luppino \& Tonry (1993) for M31, and is somewhat fainter than
Pahre \& Mould's (1994) measurement of 
$M_{K_{sh}}\,{=}\,{-}5.77\,{\pm}\,0.18$,
although the values are well within the errors.  
Our calibration
was used to compute distance moduli for the galaxies in our survey,
and we list distances in Table~\ref{fornaxm-Mtab}
with the $I$-band SBF distance moduli for comparison.
The comparison between $I$ and \Kp\ distance moduli is presented
graphically in Figure~\ref{distmodfig}.  The line shown is not a fit
to the data, but simply the line with a slope of 1 which passes through
the origin.  
The high-$S/N$ measurements are consistent with the $I$-band SBF distances.
The errors listed for \Kp\ distance moduli are the individual errors in
\mKp, and do not include the zero-point uncertainty.
Mean \Kp\ SBF distances for each cluster are included with their 
standard deviations,
where only the high-$S/N$ measurements were included in the mean.
Once again, we note that since the same Cepheid calibration 
was used for all distances tabulated, a systematic change in the 
Cepheid distance scale will affect the $I$ and \Kp\ SBF distances equally.

\begin{deluxetable}{ccccccccc}
\tablecaption{Comparison of Measured Distances \label{fornaxm-Mtab}}
\tablewidth{0pc}
\tablehead{
\colhead{Galaxy}&
\colhead{\v-i\tablenotemark{a}} &
\colhead{$\pm$} & 
\colhead{\Mg2\tablenotemark{b}} &
\colhead{$(m{-}M)_{\rm Ceph}$\tablenotemark{c}} &
\colhead{$\m-Mbar_{I}$\tablenotemark{a}} &
\colhead{$\pm$} &
\colhead{$\m-Mbar_{\Kp}$\tablenotemark{d}}&
\colhead{$\pm$}}
\startdata
Fornax  &       &       &       & 31.30 & 31.22 & 0.06 & 31.32 & 0.24 \nl
NGC~1339& 1.169 & 0.019 & 0.328 &\nodata& 31.34 & 0.14 & 31.57 & 0.15 \nl
NGC~1344& 1.156 & 0.011 & 0.290 &\nodata& 31.17 & 0.09 & 31.19 & 0.12 \nl
NGC~1379& 1.159 & 0.019 & 0.284 &\nodata& 31.30 & 0.08 & 31.05 & 0.13 \nl 
NGC~1399& 1.241 & 0.016 & 0.361 &\nodata& 31.42 & 0.14 & 31.59 & 0.16 \nl
NGC~1404& 1.238 & 0.013 & 0.344 &\nodata& 31.27 & 0.08 & 31.43 & 0.14 \nl
\tablevspace{10pt}
Eridanus &       &       &       &\nodata& 31.71 &\nodata&31.80&\nodata\nl
NGC~1395& 1.239 & 0.012 & 0.340 &\nodata& 31.63 & 0.41 & 31.04 & 0.17 \nl
NGC~1400& 1.216 & 0.009 & 0.336 &\nodata& 31.67 & 0.14 & 32.07 & 0.25 \nl
NGC~1407& 1.264 & 0.010 & 0.354 &\nodata& 32.02 & 0.19 & 31.80 & 0.15 \nl
NGC~1426& 1.176 & 0.009 & 0.303 &\nodata& 31.80 & 0.08 & 31.28 & 0.30 \nl
\tablevspace{10pt}
Virgo    &       &       &     &$31.06\pm0.13$& 30.96 & 0.05 & 31.06 & 0.12 \nl
NGC~4365& 1.223 & 0.011 & 0.348 &\nodata& 31.70 & 0.09 & 31.70 & 0.11 \nl
NGC~4406& 1.175 & 0.008 & 0.336 &\nodata& 31.19 & 0.07 & 31.06 & 0.10 \nl
NGC~4472& 1.246 & 0.011 & 0.345 &\nodata& 30.94 & 0.06 & 30.91 & 0.11 \nl
NGC~4489& 1.076 & 0.015 & 0.222 &\nodata&\nodata&\nodata&30.60 & 0.35 \nl
NGC~4552& 1.206 & 0.015 & 0.352 &\nodata& 30.86 & 0.08 & 31.05 & 0.11 \nl
NGC~4578& 1.154 & 0.015 & 0.312 &\nodata& 31.25 & 0.08 & 31.15 & 0.30 \nl
NGC~4636& 1.225 & 0.009 & 0.341 &\nodata& 30.83 & 0.08 & 31.20 & 0.15 \nl
\tablevspace{10pt}
Local Group&&&&&&&&\nl
M31& 1.240 & 0.007 & 0.334&$24.43\pm0.10$& 24.38 & 0.05 & 24.43 & 0.14 \nl
M32& 1.145 & 0.007 & 0.199 &\nodata& 24.48 & 0.05 & 24.17 & 0.14 \nl
\enddata
\tablenotetext{a}{Tonry 1997}
\tablenotetext{b}{\Mg2\ values for the Virgo cluster and local group 
galaxies are from Worthey et al. 1992 and Worthey 1993a;  for the
Fornax and Eridanus clusters, we adjusted values from Faber et al. 1989
to the Worthey scale by adding 0.027 mag.}
\tablenotetext{c}{See text for references.}
\tablenotetext{d}{\Kp\ Distance moduli for clusters are averages 
of the high-$S/N$ ratio observations only (NGC~4365 was excluded from 
the Virgo average because it resides in the background W~cloud).  
Errors for the mean cluster distances are standard deviations.
Individual distance moduli were computed using the calibration $\MKp=-5.61$. 
Uncertainties on individual distance moduli do not include the calibration
error from \MKp.  Local Group measurements were taken from Luppino \& Tonry
1993.}
\end{deluxetable}

\subsubsection{Fluctuation Magnitudes and \mg2\ Index \label{mg2section}}

\begin{figure}
\epsscale{0.8}
\plotone{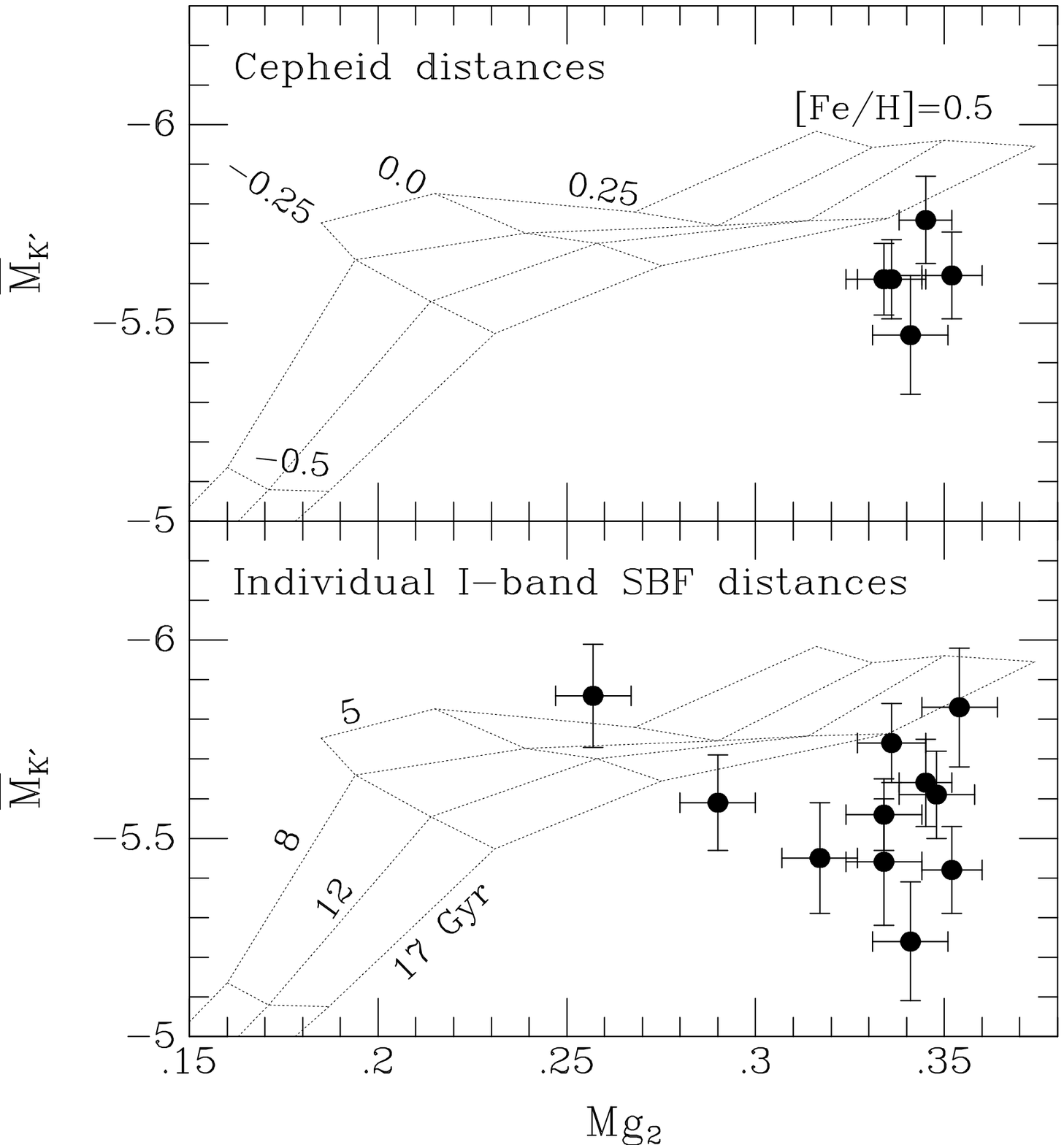}
\vspace{20pt}
\caption[Absolute Fluctuation Magnitudes as a Function of the \Mg2\ Index]
{Absolute SBF magnitudes \MKp\ are plotted versus the \Mg2\ index.
Simple stellar population models (Worthey 1997) are shown with age
and metallicity grid lines labeled.
The observations do not agree with the stellar population models,
so we cannot use the \Mg2\ index in conjunction with \Kp\ SBF magnitudes
to better understand stellar populations within the
context of the population models.
\label{Mg2}}
\end{figure}

We also compared our \Kp\ fluctuation magnitudes with Worthey's (1997)
models as as function of the \Mg2\ spectral index.
Values of the \Mg2\ index were taken from Faber et al. (1989),
Worthey et al. (1992), and Worthey (1993a).
There is a systematic offset between the Faber et al. and the
Worthey et al. measurements, the latter being larger by about 8\%.  
Seven galaxies in our sample have \Mg2\ values listed by both Worthey
et al. and by Faber et al.  The offset of 0.027 mag between the two
studies was used to adjust the Faber et al. data to the Worthey system
for the Fornax and Eridanus cluster galaxies.  
Thus we compare the Worthey (1997) models
with the most compatible \Mg2\ data.
\Mg2\ values are listed in Table~\ref{fornaxm-Mtab}.  
We plotted \MKp\ as a function of \Mg2\ for the galaxies with 
$\snsbf\,{>}\,1$ in Figure~\ref{Mg2} with the Worthey (1997) 
models superimposed, as in the previous section.  
Very few of the observations fall within the range spanned by the models.  
For our observations to be consistent with the results of the previous
section, \Mg2\ indices must be reduced by ${\sim}0.15$ mag.
The reason for the discrepancy is not known, but may be the result of
significantly enhanced [Mg/Fe] ratios (0.2 to 0.3 dex) 
observed in giant ellipticals 
compared to stars in the solar neighborhood with similar metallicities
(Worthey et al. 1992, Worthey 1994).
We do not fit \MKp\ as a function of \Mg2\ to calibrate the \Kp\ SBF
distance scale because of the small range in \Mg2\ spanned by our data
and the inconsistency with the stellar population models.

\subsubsection{Radial Gradients in SBFs\label{radialgradients}}

Radial gradients in \Kp\ fluctuation magnitudes potentially
reveal changing ages and metallicities within the stellar population
of a galaxy.
Tonry (1991) showed that the color gradient in NGC~205 correlates with
a change in the $I$-band fluctuation magnitude \mI.
A younger stellar population at the center of NGC~205 gives rise to  
brighter $I$-band fluctuations at smaller radii 
than in the redder outer regions.
Sodemann \& Thomsen (1995) discuss the implications
of a radially increasing $I$-band fluctuation amplitude observed in NGC~3379,
opposite the trend seen in NGC~205.
At $I$ the effects of age and metallicity on \mI\ are largely degenerate,
making it difficult to interpret radial changes in fluctuation magnitude.
Older stellar populations have fainter \mI\ magnitudes, but populations
with higher metallicity also have fainter \mI\ magnitudes, 
making it impossible
to distinguish between old, metal-poor and young, metal-rich populations.
At \Kp, on the other hand, the degeneracy is partially broken.  
The Worthey models (1993b, 1997) 
predict that IR fluctuations should fade with increasing age and 
decreasing metallicity, allowing us to separate young, metal-rich and
old, metal-poor populations. 
If an elliptical galaxy is formed by merging which triggers
a burst of new star formation, we would expect the outer envelope 
of such a galaxy to be older and more metal-poor than the nuclear region 
(Faber et al. 1995).  
The brighter $I$-band fluctuation magnitudes observed in the outer regions
of NGC~3379 may be consistent with this scenario 
if the change is \mI\ is principally due to a decrease in metallicity;
in NGC~205 the change in \mI\ must be due to age differences between the 
populations.
Pahre \& Mould (1994) observed no radial gradient in the 
$K$-band fluctuation magnitude in NGC~3379, but  
a modest gradient may be hidden in the 
scatter of the individual measurements.

While the prospects for using IR SBFs to reveal radial gradients
in stellar populations are good, 
the data presented here do not have a sufficiently high $S/N$
ratio to detect radial changes in fluctuation amplitude;
the uncertainties due to background subtraction and estimated
residual variance mask the modest changes in \mKp\ that
might result from differences in stellar populations within a galaxy.
If the outer regions of an elliptical galaxy are older and more metal-poor
than the center, the \Kp\ fluctuation amplitude will decrease with radius.
In our data, uncorrected fluctuation amplitudes usually
increase with radius, sometimes changing by a factor of two.
Even if we could imagine a scenario that creates a galaxy with the 
older stars at the center, the variations in stellar population that would
be required to explain such large changes in fluctuation
amplitude are not plausible. 
Instead, we interpret the observed SBF gradients in our data 
to be indicators of residual variations in the sky background.

Using IR SBFs to reach definitive conclusions regarding stellar population
gradients in nearby galaxies will require high-$S/N$ ratio
data in which corrections for residual variance are negligible.
Although we have not conclusively measured IR SBF gradients 
it should be emphasized
that the breaking of the age and metallicity degeneracy potentially 
makes IR SBFs
a powerful probe for distinguishing old, metal-poor populations from 
younger, metal-rich stars. 
For example, the range in \v-i\ in NGC~205 should give rise to a change 
in \mKp\ of ${\sim}0.5$ mag, with \mKp\ becoming fainter with radius.  
If an IR SBF gradient were detected
in NGC~3379, we could explain the very different optical SBF properties
of these two galaxies in terms of the ages and metallicities of the stellar
populations. 

\section{Summary}

We have completed a \Kp\ (2.1~\micron) survey of surface brightness
fluctuations in 16 giant elliptical galaxies in the Virgo, Fornax, and 
Eridanus clusters.  We also observed M31 and M32 in the $J$ and $H$ 
bands.
From this study we draw the following conclusions:

$K$-band SBFs can be measured reliably and used to determine 
distances to early-type galaxies.  Because SBFs are relatively bright in the
IR and the seeing is typically very good, integration times can
be quite modest.  However, observations must be sufficiently deep
to adequately sample the GC luminosity function and remove sources
of residual variance.  Low-$S/N$ ratio measurements are unreliable.
Optical images can be used to identify and
remove GCs and background galaxies that are not detected in the IR,
improving the IR SBF measurement.  
We empirically calibrated the \Kp\ SBF distance scale 
using Cepheid distances to M31 and the Virgo cluster.
The absolute \Kp\ fluctuation magnitude
is $\MKp\,{=}\,{-}5.61\,{\pm}\,0.06$ (statistical error) with a total
uncertainty of 0.12 mag. 
No significant change in \MKp\ with \v-i\ was observed over the range in 
color spanned by the galaxies in this sample.  
Accurate color measurements are not required to measure the \Kp\ SBF 
distance to a galaxy.  

\Kp\ SBF magnitudes are consistent with predictions from simple stellar 
population models.  
The lack of a correlation in \MKp\ with \v-i\ is best explained by a spread
in ages among the galaxies observed.  The redder ellipticals are
consistent with 12 to 17 Gyr stellar population models, while the bluer
galaxies in our sample must have younger 5 to 8 Gyr populations.  
Metallicities appear
to vary less than ages, with the typical galaxy having [Fe/H]\,=\,${-}0.25$. 
The stellar population models show that the age-metallicity
degeneracy is broken with $K$-band SBFs,
allowing one to distinguish between old, metal-poor 
and young, metal-rich populations. 
Examining the radial variation in IR SBFs will help distinguish between
different galaxy formation scenarios.
Our observations do not agree with the
relationship between \MKp\ and the \Mg2\ 
index predicted by the stellar population models.

\acknowledgments
We are especially indebted to Guy Worthey for providing 
the custom \Kp\ stellar population models and the $K$-band K corrections.
Many sections of this paper were strengthened by the addition of his
models.  We also thank John Tonry and his collaborators for allowing us
to use $I$-band images from their SBF survey to identify and remove
globular clusters and background galaxies from our \Kp\ images.
We gratefully acknowledge helpful conversations with Michael Pahre and 
Michael Liu.
We acknowledge the support of grants NSF AST9401519 and 
STScI GO-06579.01-95A.
This research has made use of the NASA/IPAC Extragalactic Database (NED)   
which is operated by the Jet Propulsion Laboratory, California Institute   
of Technology, under contract with the National Aeronautics and Space      
Administration.                                                            



\clearpage
\begin{thebibliography}{}

\bibitem[]{}
Ajhar, E. A. \& Tonry, J. T. 1994, \apj, 429, 557

\bibitem[]{}
Blakeslee, J. P. 1997, \apjl, 481, L59

\bibitem[]{}
Blakeslee, J. P. \& Tonry, J. L. 1995, \apj, 442, 579

\bibitem[]{}
Blakeslee, J. P. \& Tonry, J. L. 1996, \apjl, 465, L19

\bibitem[]{}
Blakeslee, J. P., Tonry, J. L., \& Metzger, M. R. 1997, \aj, 114, 482

\bibitem[]{}
Burstein, D. \& Heiles, C. 1984, \apjs, 54, 33

\bibitem[]{}
Buzzoni, A. 1995, \apjs, 98, 69

\bibitem[]{}
Casali, M. \& Hawarden, T. G. 1992, JCMT-UKIRT Newsletter, 4, 33

\bibitem[]{}
Cohen, J. G., Frogel J. A., Persson, S. E., \& Elias, J. H. 1981, 
\apj, 249, 481

\bibitem[]{}
Cowie, L. L., Gardner, J. P., Hu, E. M., Songaila, A., Hodapp, K.-W., \&
Wainscoat, R. J. 1994, \apj, 434, 114

\bibitem[]{}
Djorgovski, S. et al., 1995, \apjl, 438, 13

\bibitem[]{}
Faber, S. M., Trager, S. C., Gonzalez, J. J., \& Worthey, G. 1995,
in Stellar Populations, I.A.U. Symp. 164, eds. P. C. van der Kruit \&
G. Gilmore (Kluwer, Dordrecht), p. 249

\bibitem[]{}
Faber, S. M., Wegner, G., Burstein, D., Davies, R. L., Dressler, A.,
Lynden-Bell, D., \& Terlevich, R. J. 1989, \apjs, 69, 763

\bibitem[]{}
Feast, M. W. \& Catchpole, R. M. 1997, \mnras, 286, L1

\bibitem[]{}
Ferrarese, L. et al. 1996, \apj, 464, 568

\bibitem[]{}
Forbes, D. 1996, \aj, 112, 954 

\bibitem[]{}
Freedman, W. L. \& Madore, B. F. 1990, \apj, 365, 186

\bibitem[]{}
Freedman, W. L. et al. 1994, \nat, 371, 757

\bibitem[]{}
Frogel, J. A., Persson, S. E.,  Aaronson, M., \& Matthews, K. 1978, 
\apj, 220, 75

\bibitem[]{}
Glass, I. S. 1984, \mnras, 211, 461

\bibitem[]{}
Hodapp, K.-W. et al. 1996, New Astronomy, 1, 177

\bibitem[]{}
Hodapp, K.-W., Rayner, J., \& Irwin, E. 1992, \pasp, 104, 441

\bibitem[]{}
Jacoby, G. H., Branch, D., Ciardullo, R., Davies, R. L.,
Harris, W. E., Pierce, M. J., Pritchet, C. J., Tonry, J. L.,
\& Welch, D. L. 1992, \pasp, 104, 599

\bibitem[]{}
Jensen, J. B., Luppino, G. A., \& Tonry, J. L. 1996, \apj, 468, 519 
(JLT)

\bibitem[]{}
Luppino, G. A. \& Tonry, J. L. 1993, \apj, 410, 81

\bibitem[]{}
Madore, B. F. \& Freedman, W. L. 1997, \apj, 492, 110

\bibitem[]{}
Pahre, M. A. \& Mould, J. R. 1994, \apj, 433, 567

\bibitem[]{}
Persson, S. E., Cohen, J. G., Sellgren, K., Mould, J., \& Frogel, J. A. 1980
\apj, 240, 779

\bibitem[]{}
Persson, S. E., Frogel, J. A., \& Aaronson, M. 1979, \apjs, 39, 61

\bibitem[]{}
Pierce, M. J., Welch, D. L., McClure, R. D., van den Bergh, S., 
Racine, R., \& Stetson, P. B. 1994, \nat, 371, 385


\bibitem[]{}
Saha, A., Sandage, A., Labhardt, L, Tammann, G. A., Macchetto, F. D., \&
Panagia, N. 1996a, \apj, 466, 55

\bibitem[]{}
Saha, A., Sandage, A., Labhardt, L, Tammann, G. A., Macchetto, F. D., \&
Panagia, N. 1996b, \apjs, 107, 693

\bibitem[]{}
Sandage, A., Saha, A., Tammann, G. A., Labhardt, L., Panagia, N., \&
Macchetto, F. D. 1996, \apjl, 460, L15

\bibitem[]{}
Schechter, P. L., Mateo, M., \& Saha, A. 1993, \pasp, 105, 1342 

\bibitem[]{}
Secker, J. \& Harris, W. E. 1993, \aj, 105, 1358

\bibitem[]{}
Silbermann, N. 1998, private communication

\bibitem[]{}
Sodemann, M. \& Thomsen, B. 1995, \aj, 110, 179

\bibitem[]{}
Sodemann, M. \& Thomsen, B. 1996, \aj, 111, 208

\bibitem[]{}
Tonry, J. L. 1991, \apjl, 373, L1

\bibitem[]{}
Tonry, J. L. 1997, private communication

\bibitem[]{}
Tonry, J. L., Ajhar, E. A., \& Luppino, G. A. 1990, \aj, 100, 1416

\bibitem[]{}
Tonry, J. L., Blakeslee, J. P., Ajhar, E. A., \& Dressler, A. 1997, 
\apj, 475, 399

\bibitem[]{}
Tonry, J. L. \& Schneider, D. P. 1988, \aj, 96, 807

\bibitem[]{}
Wainscoat, R., Cohen, M., Volk, K., Walker, H. J., \& Schwartz, D. E. 1992,
\apjs, 83, 111

\bibitem[]{}
Wainscoat, R. \& Cowie, L. 1992, \aj, 101, 332

\bibitem[]{}
Worthey, G. 1993a, \apj, 409, 530

\bibitem[]{}
Worthey, G. 1993b, \apj, 418, 947

\bibitem[]{}
Worthey, G. 1994, \apjs, 95, 107

\bibitem[]{}
Worthey, G. 1997, private communication

\bibitem[]{}
Worthey, G., Faber, S. M., \& Gonzalez, J. J. 1992, \apj, 398, 69

\end{thebibliography}
\end{document}